\documentclass[aps,prl,longbibliography,showpacs,twocolumn,superscriptaddress,amsmath,amssymb,verbatim]{revtex4-2}
\usepackage[T1]{fontenc}
\usepackage[latin9]{inputenc}
\usepackage{bigints}
\usepackage[usenames,dvipsnames]{color}
\usepackage{float}
\usepackage{amssymb,amsmath}
\usepackage{graphicx}
\usepackage{epsfig}
\pdfoutput=1
\usepackage[normalem]{ulem}
\usepackage{booktabs}
\usepackage{setspace}
\usepackage{esint}
\usepackage{natbib}
\usepackage{bigints}
\usepackage{color}
\usepackage{babel}
\usepackage{amsmath,amsfonts,amssymb}
\usepackage{pifont}
\usepackage{rotating}
\usepackage{graphicx}    
\usepackage{epstopdf}
\usepackage{color}
\usepackage[caption=false]{subfig}
\usepackage{soul}
\usepackage{color}
\usepackage{babel}
\usepackage{array}
\usepackage{float}
\usepackage{booktabs}
\usepackage{setspace}
\usepackage{esint}
\usepackage{chemformula}
\usepackage[unicode=true,pdfusetitle,
bookmarks=true,bookmarksnumbered=true,bookmarksopen=false,
breaklinks=true,pdfborder={0 0 0},pdfborderstyle={},backref=false,colorlinks=true]
{hyperref}
\hypersetup{
	linkcolor=red, citecolor=blue,  urlcolor=blue}
\definecolor{dark-blue}{rgb}{0.15,0.15,0.4}

\usepackage{fancyhdr} 
\makeatletter
\@ifundefined{textcolor}{}
{%
 \definecolor{BLACK}{gray}{0}
 \definecolor{WHITE}{gray}{1}
 \definecolor{RED}{rgb}{1,0,0}
 \definecolor{GREEN}{rgb}{0,1,0}
 \definecolor{BLUE}{rgb}{0,0,1}
 \definecolor{CYAN}{cmyk}{1,0,0,0}
 \definecolor{MAGENTA}{cmyk}{0,1,0,0}
 \definecolor{YELLOW}{cmyk}{0,0,1,0}
}

\newcommand{\SPHIDE}[1]{{}}

\makeatother

\usepackage{babel}
\begin{document}

\title{   \ch{(H,Li)_{6}Ru_{2}O_{6}} : a possible zero-field Ru$^{3+}$-based Kitaev Quantum Spin Liquid}

	\author{Sanjay Bachhar} 
	\email{sanjayphysics95@gmail.com}
	\affiliation{Department of Physics, Indian Institute of Technology Bombay, Powai, Mumbai 400076, India}

	\author{M. Baenitz}
	\affiliation{Max Planck Institute for Chemical Physics of Solids, 01187 Dresden, Germany}

	\author{Hubertus
		Luetkens}
    \affiliation{Laboratory for Muon Spin Spectroscopy, Paul Scherrer Institute, CH-5232 Villigen PSI, Switzerland}
    
   \author{John Wilkinson}
   \affiliation{ISIS Pulsed Neutron and Muon Source, STFC Rutherford Appleton Laboratory,
   	Harwell Campus, Didcot, Oxfordshire OX110QX, United Kingdom}
		\author{Sumiran Pujari}
	\affiliation{Department of Physics, Indian Institute of Technology Bombay, Powai, Mumbai 400076, India}

	\author{A.V. Mahajan}
	\email{mahajan@phy.iitb.ac.in}
	\affiliation{Department of Physics, Indian Institute of Technology Bombay, Powai, Mumbai 400076, India}

\begin{abstract}

We report the synthesis and properties of \ch{(H,Li)_{6}Ru_{2}O_{6}}, which is shown to be
a $J_{eff}=\frac{1}{2}$ system made out of Ru$^{3+}$ moments in a honeycomb geometry.
Bulk magnetization, heat capacity, nuclear magnetic resonance (NMR), and muon spin relaxation ($\mu$SR) 
rule out the presence of static moments or any spin glass phase down to 84 mK. 
All techniques suggest a crossover to a liquid-like state below about 40 K. 
The $^{7}$Li nuclear magnetic resonance (NMR) shift data suggest a non-zero $T$-independent 
spin susceptibility at low $T$. 
In zero field, $C_m/T$ shows $T^{-0.9}$ divergence which is consistent with vacancy-induced effects on low-energy
excitations of the pristine Kitaev spin liquid.
With field, power-law variations 
in the $^{7}$Li NMR spin-lattice relaxation rate 1/T$_{1}$ and 
magnetic heat capacity $C_{m}$ show quantitatively new scaling behaviors.
A two-step entropy release in heat capacity is also observed  putatively
from $Z_{2}$ flux (low-$T$ step) and itinerant Majorana fermions (high-$T$ step). 
Based on these 
findings, we propose that \ch{(H,Li)_{6}Ru_{2}O_{6}} realizes a Kitaev spin liquid with
no evidence of inherent magnetic ordering in zero field unlike $\alpha$-\ch{RuCl3}
where an 8 Tesla field is required to suppress magnetic order.


\end{abstract}

\maketitle


\textit{Introduction.$-$} Kitaev quantum spin liquids (KQSL) continue to evoke interest due to their novel and exotic ground state and excited state properties~\cite{Kitaev2006,Takagi_review_2019, Perkins2021,Trebst2019, Perkins2021_Dec}. They are expected to harbor itinerant Majorana fermion excitations which, apart from being intrinsically interesting, are thought to be useful for quantum computing. An example of a promising Kitaev material is the layered, honeycomb material $\alpha$-\ch{RuCl3} \cite{BaeK2017,Arnab2017,Baenitz2015}  
which however undergoes magnetic ordering around 7 K and requires fields greater than about 
80 kOe \cite{BaeK2017} to reveal KQSL features.  It has been a challenge to find a material with KQSL physics in absence of magnetic field in general including Ru-based compounds. $\alpha$-\ch{Li2IrO3} is another promising KQSL candidate material which again shows AFM ordering below $T_{N} \sim 15$ K \cite{Williams2016}. 
Takagi {\it{et al.}} \cite{Kitagawa2018} showed that hydrogen intercalation in $\alpha$-\ch{Li2IrO3} results in the formation of \ch{H3LiIr2O6} which shows KQSL behaviour with novel low-energy excitations.   Given the requirement of a large spin-orbit coupling for Kitaev materials, 4$d$/5$d$-based honeycomb lattice systems are good starting points to look for new Kitaev systems, for instance moments from Ir$^{4+}$/Ru$^{3+}$ with $d^{5}$ configuration and $J_{\text{eff}}=\frac{1}{2}$ moments~\cite{Kim_etal_PRL_2008, Takagi_review_2019, Trebst_review_2022,Zwartsenberg2020,Razpopov2023}.
In the iridate case, longer intercalation times starting from $\alpha$-\ch{Li2IrO3} gives \ch{H5LiIr2O6}~\cite{Schenghai2020} which however contains Ir$^{3+}$ (5$d^{4}$, $J_{\text{eff}} = 0$). There are relatively fewer compounds with Ru$^{3+}$ \cite{Bhattacharyya2023,Razpopov2023} with $\alpha$-\ch{RuCl3} 
being the only honeycomb example.
With this background, we set out to prepare \ch{H5LiRu2O6} (\ch{(H,Li)_{6}Ru_{2}O_{6}}) by intercalating excess H in \ch{Li2RuO3} as a possible Kitaev material.
We obtained \ch{(H,Li)_{6}Ru_{2}O_{6}} with a honeycomb network of Ru$^{3+}$ as in $\alpha$-\ch{RuCl3}. However, unlike $\alpha$-\ch{RuCl3}, \ch{(H,Li)_{6}Ru_{2}O_{6}} intriguingly shows no sign of magnetic order down to about 500 mK. Our measurements pertaining to the magnetism of this new compound find the emergence of a KQSL state at temperatures below 40 K with gapless excitations.

\begin{figure}[h]
	\centering{}\includegraphics[width=0.85\linewidth]{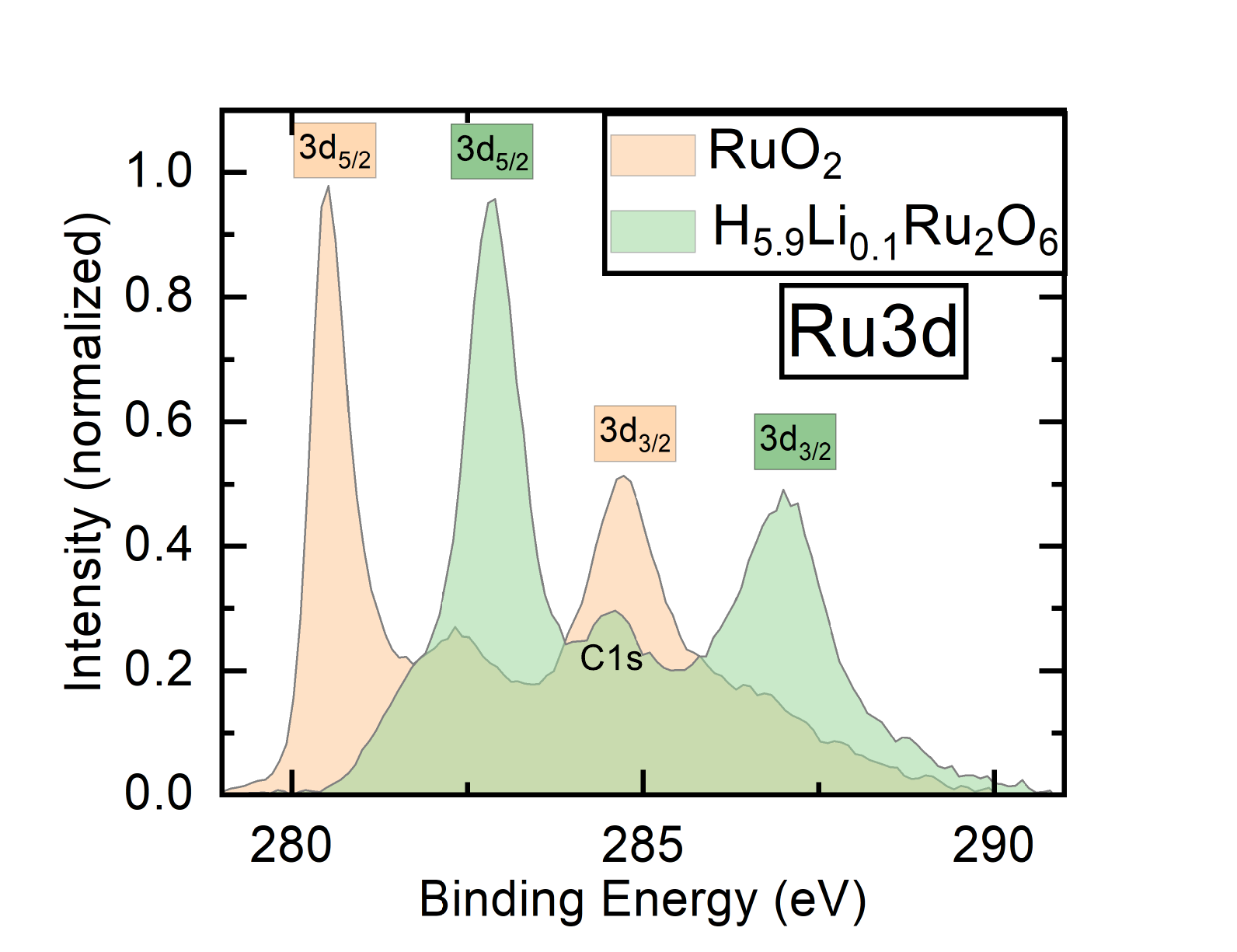}\caption{\label{XPS}{\small{} Ru3d XPS Spectra of \ch{H_{5.9}Li_{0.1}Ru2O6}} where binding energies of 3d$_{5/2}$ (282.9 eV) and 3d$_{3/2}$ (287 eV) are similar to $\alpha$-\ch{RuCl3} (Ru is in +3 oxidation state)\cite{Morgan2015}. Also, Ru3d XPS spectra of \ch{RuO2} (Ru is in +4 oxidation state) is shown for comparison and binding energies are consistent with literature \cite{Morgan2015}. C(1s) reference position was taken 284.7 eV for both materials.}
\end{figure}

  Before going into the details of the measurements, we discuss some relevant details
about the structure of the compound.
Previously, silver intercalation in \ch{Li2RuO3} led to \ch{Ag3LiRu2O6} with Ru$^{4+}$ \cite{RKumar2019}. That compound crystallizes in $C2/m$ space group and there was no sign of dimerization unlike \ch{Li2RuO3} \cite{Kimber2010,Takagi2022,RKumar2019}. \ch{Ag3LiRu2O6} is possibly an example of honeycomb quantum magnet of Khaliullin type (excitonic magnetism in 4$d^{4}$ based system)\cite{Takagi2022,Khaliullin2013_prl,RKumar2019,Khaliullin2019_prl,Khaliullin2019_prb}. Instead of the heavier Ag, a lighter and smaller H can be intercalated in excess amount as observed in \ch{H5LiIr2O6}. Excess H-intercalation in \ch{Li2RuO3} leads to the replacement of all the inter-layer Li atoms by H atoms and also of a fraction of the in-plane Li atoms. There is a contraction of the $c$-parameter of the lattice (from 5.13 {\AA} to 5.01 {\AA}) due to the lighter and smaller H atoms replacing the inter-layer Li atoms (see the supplementary materials (SM) \cite{SM}). The $c$-axis shortening does tune the magnetism as observed in \ch{H3LiIr2O6}, a KQSL (\emph{c}-parameter reduced from 5.12 {\AA}  to 4.87 {\AA} after H-intercalation in \ch{Li2IrO3}). The structural analysis of HLRO reveals a nearly perfect honeycomb network with a Ru-Ru bond distance of $\sim$ 2.96 {\AA} and a $d-p-d$ bond angle of $\sim$ 95{\textdegree}. The presence of Ru$^{3+}$ ($J_{\text{eff}}=1/2$) is inferred from x-ray photoelectron spectroscopy (XPS) analysis. See Fig.~\ref{XPS} (see SM \cite{SM} for further details). 
Also, an entropy recovery of $R\ln2$ (by about 100 K) is seen in our specific heat data
consistent with $J_{\text{eff}}=\frac{1}{2}$ magnetic moments. 
Li-quantification through $^{7}$Li NMR Spectral analysis (see SM \cite{SM}) in \ch{(H,Li)_{6}Ru_{2}O_{6}} led us to the 
chemical formula, \ch{H_{5.9}Li_{0.1}Ru2O6}~\cite{SM} to be referred to as HLRO
in the rest of the paper.  

Our main findings on the magnetism of HLRO are as follows: 
Bulk magnetization, heat capacity, NMR, $\mu$SR rule out the presence of static moments 
or any spin glass phase down to 500 mK. 
There is a crossover to a liquid-like state below about 40 K.
The zero-field divergence of the magnetic heat capacity divided by temperature ($\frac{C_m}{T}$)
is consistent with a KQSL scenario in presence of vacancies.
$\frac{C_m}{T}$ also shows scaling with increasing field which has been phenomenologically
modeled.
$^{7}$Li NMR shift is non-zero and  $T$-independent indicating a Pauli-like spin susceptibility at low-T 
as seen in many QSLs. NMR spin-lattice relaxation rates show a power-law variation consistent
with gapless excitations. Finally, a two-step mangetic entropy release is observed 
suggestive of Z$_{2}$ flux excitations at low-$T$ and itinerant Majorana fermions at high-$T$.

\begin{figure}[h]
	\centering{}\includegraphics[width=0.85\linewidth]{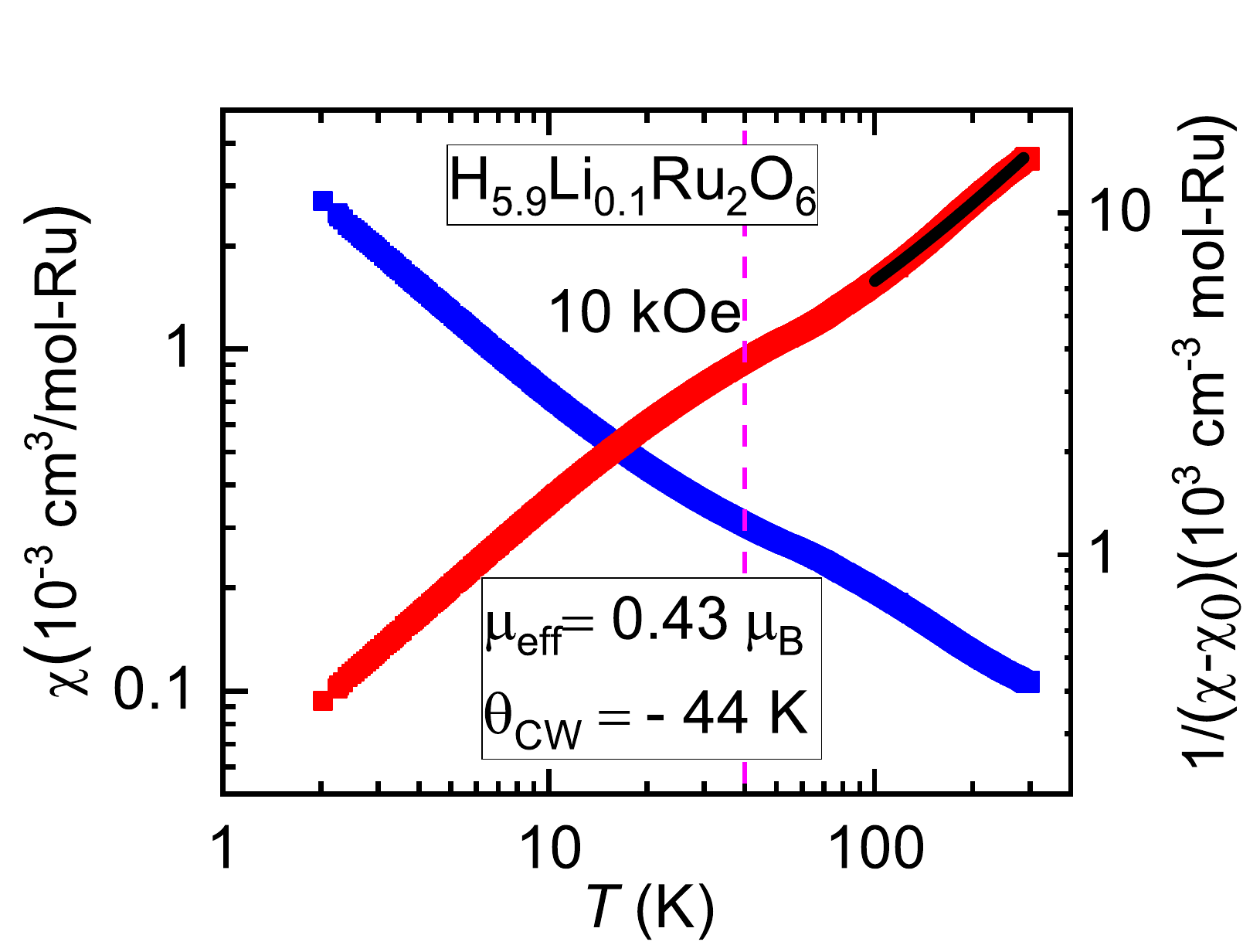}\caption{\label{HLRO_96h_chi}{\small{}(a) The left $y$-axis shows the temperature dependence of the susceptibility, $\chi(T)\equiv$$\frac{M(T)}{H}$ (blue squares) of \ch{H_{5.9}Li_{0.1}Ru_{2}O_{6}} measured in a field of 10 kOe in zero field cooled (ZFC) condition and the right $y$-axis shows the inverse susceptibility (red squares) free from temperature independent susceptibility, $\chi$$_{0}$. The black solid line on red square symbols of 1/($\chi-\chi_{0}$) shows the Curie-Weiss fit in the temperature range 100-300 K.}}
\end{figure}

\textit{Results and discussion.$-$} 
The bulk magnetic susceptibility of HLRO is shown in Fig.~\ref{HLRO_96h_chi}. It shows a Curie-Weiss like increase with decreasing temperature and a crossover around 40 K below which the susceptibility continues to increase ($ \chi_{\text{low-$T$}} \sim T^{-\eta}$ with $\eta \sim 0.9$). 
No sharp anomaly suggestive of absence of long-range order is observed down to 2 K. A Curie-Weiss fit ($\chi$= $\chi$$_{0}$ + $\frac{C}{T-\theta_{CW}}$) in the temperature range 100-300 K gives $\chi_{0} = 3.69 \times 10^{-5}$ cm$^{3}$/mol-Ru, $\theta_{CW}$ $\sim -44$ K and effective moment $\mu_{eff}$ $\sim$ 0.43 $\mu_B$. 
Absence of a bifurcation in the magnetic susceptibility between zero field cooled (ZFC) and field cooled (FC) mode in a small applied field of $H=50$ Oe in the temperature range 100 K to 2 K rules out the possibility of static/glassy moments present in the system (see SM \cite{SM}).

\begin{figure}[h]
	\centering{}\includegraphics[scale=0.33]{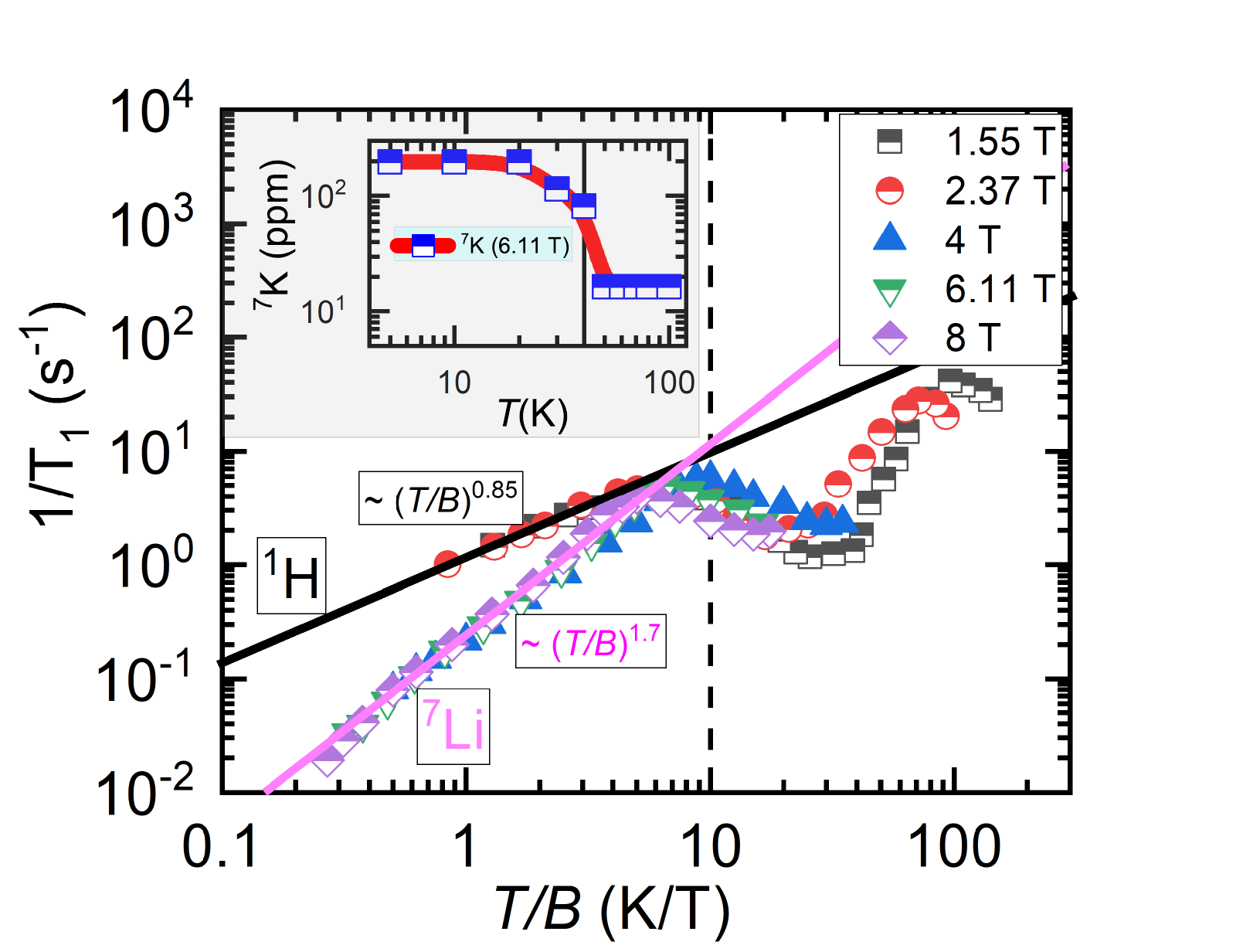}\caption{\label{T1scaling_HLRO96h}{\small{}The temperature-field scaling of the $^{7}$Li NMR spin-lattice relaxation rate, 1/$^7$T$_{1}$, with exponent $\sim$ 1.7 and $^{1}$H NMR spin-lattice relaxation rate, 1/$^1$T$_{1}$, with exponent $\sim$ 0.85 for \ch{H_{5.9}Li_{0.1}Ru_{2}O_{6}}. The data for $^{1}$H are multiplied by 13.24 for better presentation. (Inset) $^{7}$Li-NMR Shift, $^{7}$K with $T$ with crossover around 40 K (solid vertical line). }}
\end{figure}


 Next, we present results of NMR which is a powerful local probe of intrinsic spin susceptibility and low-energy excitations. The $^{7}$Li NMR shift ($^{7}$K) as function of temperature is shown in Fig.~\ref{T1scaling_HLRO96h} inset. There is a crossover around $T \sim 40$ K similar to that in bulk susceptibility (Fig.~\ref{HLRO_96h_chi}) and then it levels off down to the lowest temperature. Such $T$-independent variation has been seen in many QSLs. We have also measured $^{1}$H-NMR. No NMR-shift was observed for $^{1}$H down to the lowest temperature similar to \ch{H3LiIr2O6} possibly due to weak hyper-fine coupling. The variation of  $^{1}$H and $^7$Li NMR 1/T$_1$ as a function of temperature for  HLRO in various applied magnetic fields is shown in Fig.~\ref{T1scaling_HLRO96h}. A crossover is again seen at $\sim$ 40 K similar to other measurements. Below 40 K, 1/T$_{1}$ exhibits a power law variation with temperature. While the crossover at $\sim$ 40 K is not due to long-range ordering or glassy magnetism, the power-law variation below 40K indicates gapless excitations. 
Furthermore, 1/T$_{1}$ scales with $\frac{T}{B}$ with an exponent $\sim$ 1.7 for $^{7}$Li and 
0.85 for $^{1}$H (Fig.~\ref{T1scaling_HLRO96h}) that can be phenomenologically attributed to 
field-dependent density of states~\cite{Kitagawa2018}. 


\begin{figure}[h]
	\centering{}\includegraphics[scale=0.35]{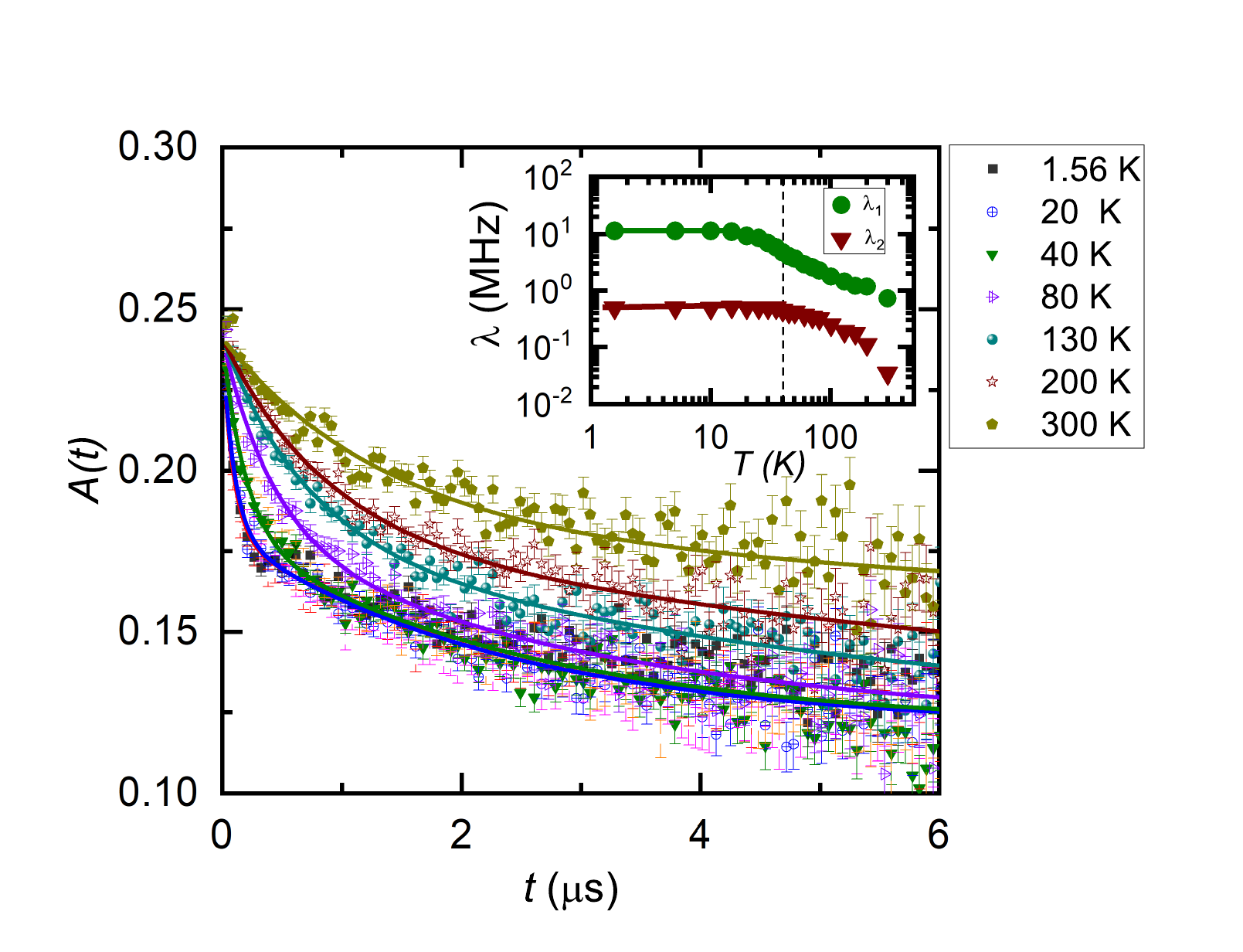}\caption{\label{Lambda_T}{\small{}Muon asymmetry as a function of time in a 100 Oe longitudinal field in the temperature range: 1.56 K-300 K. (Inset)  Muon relaxation rate $\lambda$ as a function of temperature. Two components of relaxation, $\lambda_1$ (olive circles) and $\lambda_2$ (wine triangles) with A$_{1}$/A$_{2}$ $\sim$ 1 are observed.}}
\end{figure}


Another piece of evidence we now present is that of local moment dynamics in HLRO through $\mu$SR measurements which were carried out at the GPS facility in PSI, Switzerland and dilute fridge measurement at ISIS,UK. Zero field (ZF) muon depolarisation data~\cite{SM} show an exponential decay with no hint of oscillations down to 84 mK. This indicates the  absence of long range ordering consistent with our other measurements. The absence of $1/3$-tail in muon asymmetry signals that the moments remain dynamic throughout. Since ZF measurements include a contribution from nuclear moments, we have also taken data in a longitudinal field (LF) of 100 Oe to remove the depolarisation due to nuclear moments. This is shown in Fig.~\ref{Lambda_T}. 
We could fit this purely electronic contribution to the muon relaxation asymmetry well using $A(t)=A_{1}e^{-\lambda_{1}t} + A_{2}e^{-\lambda_{2}t} + A_{3}$~\cite{SM}.
The two components of relaxation ($\lambda_1$ and $\lambda_2$) occur with nearly equal relative weights (A$_{1}$/A$_{2}$ $\sim$ 1) independent of temperature as shown in the inset of Fig.~\ref{Lambda_T}.  This is likely due to the presence of two muon stopping sites \cite{AVM2021}. The muon relaxation rate increases gradually with decreasing temperature and finally levels off below about 40 K indicating that the spins remains dynamic down to low-$T$. These results are similar to those of H$_{3}$LiIr$_{2}$O$_{6}$~\cite{Yang2024}. Note that the flattening of $\lambda$ in HLRO sets in at 10 times the temperature in H$_{3}$LiIr$_{2}$O$_{6}$ allowing for a larger temperature range to probe KQSL dynamics. 
 We have also monitored muon relaxation in HLRO in  longitudinal fields and find that even in a field of 3200 Oe, residual relaxation is still present (see SM \cite{SM}). The field dependency of $\lambda$ indicates the presence of long time spin correlations with a local moment fluctuation frequency about 32 MHz. This is remarkably similar to other quantum spin liquids for instance, \ch{Sr3CuSb2O9} \cite{SKundu2020} and \ch{YbGaMgO4} \cite{YMGO}. This is notably the first such observation in a KQSL among the reported KQSL like $\alpha$-\ch{RuCl3} and \ch{H3LiIr2O6}.

\begin{figure}[h]
	\centering{}\includegraphics[scale=0.35]{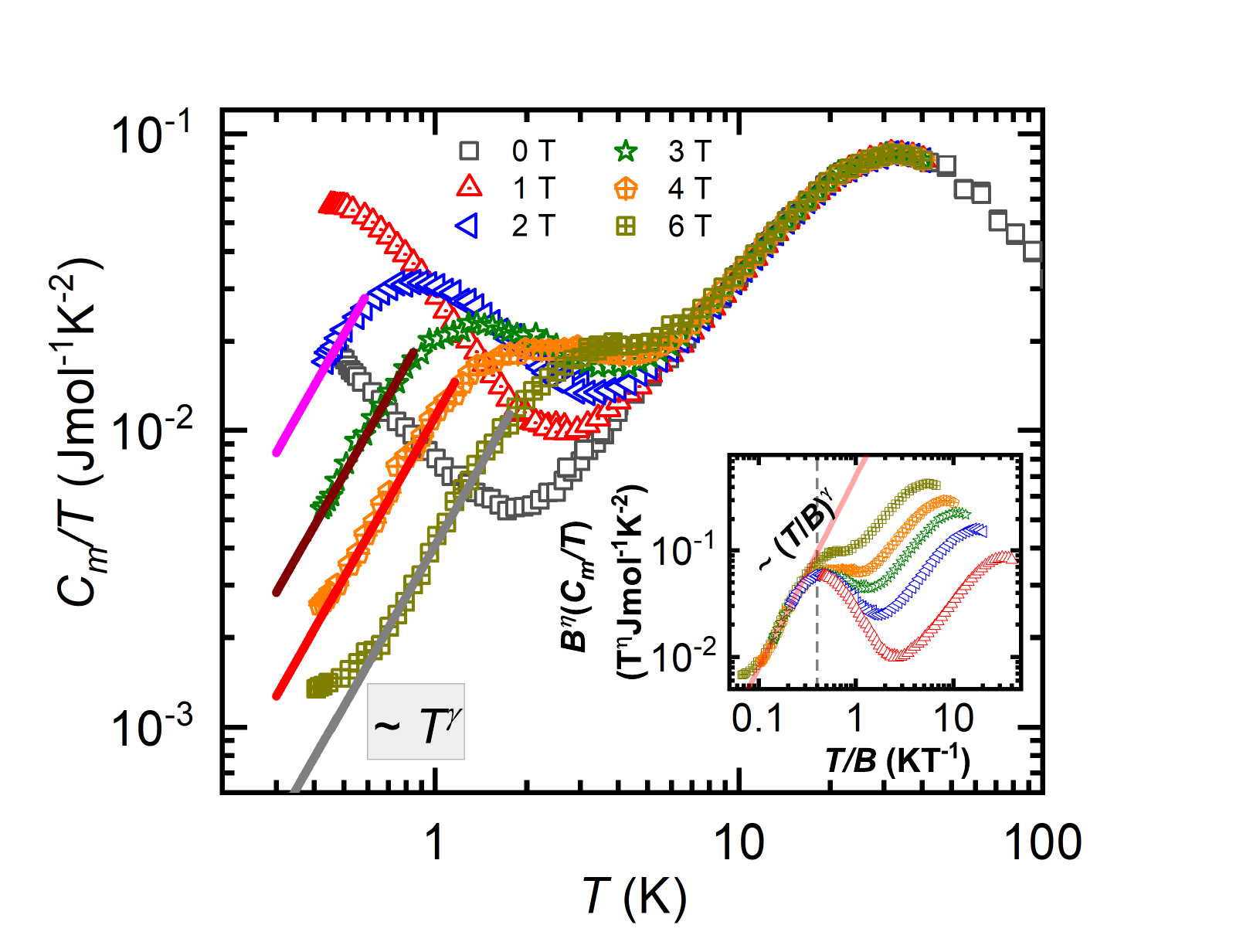}
\caption{\label{Cm_scaling}{\small{} Magnetic specific heat, $C_{m}$/$T$ versus $T$ for various fields $B$ (0-6 Tesla). (Inset)A scaling plot: $B^\eta(C_{m}/T$) versus $T/B$ with exponent,$\gamma$ $\sim$ 1.8 and $\eta \sim 0.9$ for \ch{H_{5.9}Li_{0.1}Ru_{2}O_{6}}.}}

\end{figure}

Finally, we discuss the heat capacity 
$C_{p}(T)$ of HLRO. 
The magnetic piece ($C_{m}$) is obtained by subtracting 
the $C_{p}$ of a non-magnetic analog~\cite{SM}. 
In zero magnetic field, no sharp anomaly is seen in $C_m$ down to 500 mK in agreement with our earlier conclusion of no long-range magnetic order. 
Furthermore, a non-zero $C_{m}$ was seen down to the lowest temperature measured  
indicating the absence of a spin gap. 
Fig.~\ref{Cm_scaling} shows $C_m/T$ as a function of $T$. 
In zero magnetic field, $C_{m}/T$ $\propto$ $T^{-\eta}$ ($\eta \sim 0.9$) below about 2 K. 
A rationale for such a variation has been theoretically proposed
in Ref.~\cite{Perkins2021} to arise from 
small amounts of vacancies and quasi-vacancies 
in a KQSL.
The picture is that the ground state has bound $Z_2$-flux in
low fields in the presence of vacancies. The zero-flux KQSL
is obtained upon the application of magnetic field.
With the application of a magnetic field above 1 T, 
$C_{m}/T$ follows a power-law scaling as $(T/B)^\gamma$
as shown in the inset of Fig.~\ref{Cm_scaling}
with $\gamma \sim1.8$.
This is consistent with gapless excitations
which we ascribe to the the zero-flux KQSL and weak localization of Majorana fermions \cite{Perkins2021, Perkins2021_Dec}.

We can account for the above using a phenomenological model
for the low energy excitations following Ref.~\cite{Kitagawa2018}.
An energy-symmetric fermionic density of states or DOS $D(E,B)$ of the
form $D(E,0)= \Gamma E^{-\eta}$
 gives rise to  
$C_{m}/T \propto T^{-\eta}$. $\Gamma$ is a constant. In our system, $\eta \sim 0.9$ in zero field as mentioned earlier. For comparison, $\eta$ was found to be around 0.5 in the system of Ref.~\cite{Kitagawa2018}. Note these different specific heat scaling behaviors are also accommodated in the theory of Ref.~\cite{Perkins2021}; see Fig.~13a of this paper for instance. In presence of a field,
the DOS in the low-energy region below $\alpha \mu_{B} B$ is written as $D(E, B) = \Gamma |E|^{\gamma} /(\alpha \mu_{B}B)^{\gamma+\eta}$ where $\alpha$ is some dimensionless constant.
This models a suppression of DOS at low energies.
Such a DOS leads to $B^\eta \left(\frac{C_m}{T} \right) \propto \left(\frac{T}{B} \right)^\gamma$ 
which can explain the observed scaling of $C_m/T$ as shown
in the inset of Fig.~\ref{Cm_scaling}. 
An alternative approach~\cite{Andrade2022} based on a physical picture similar to Ref.~\cite{Perkins2021} also predicts
$D(E,0) \propto E^{-\eta}$ thus giving the same zero-field power laws
as before. 
In presence of field, $C_m$ is argued to scale $\propto B^{-3\eta}T^{1+2\eta}$ in this approach.
This is also consistent with our experimental observations with $\eta \sim 0.9$.
For details, see Ref.~\cite{SM}. 



\begin{figure}[h]
	\centering{}\includegraphics[width=0.85\linewidth]{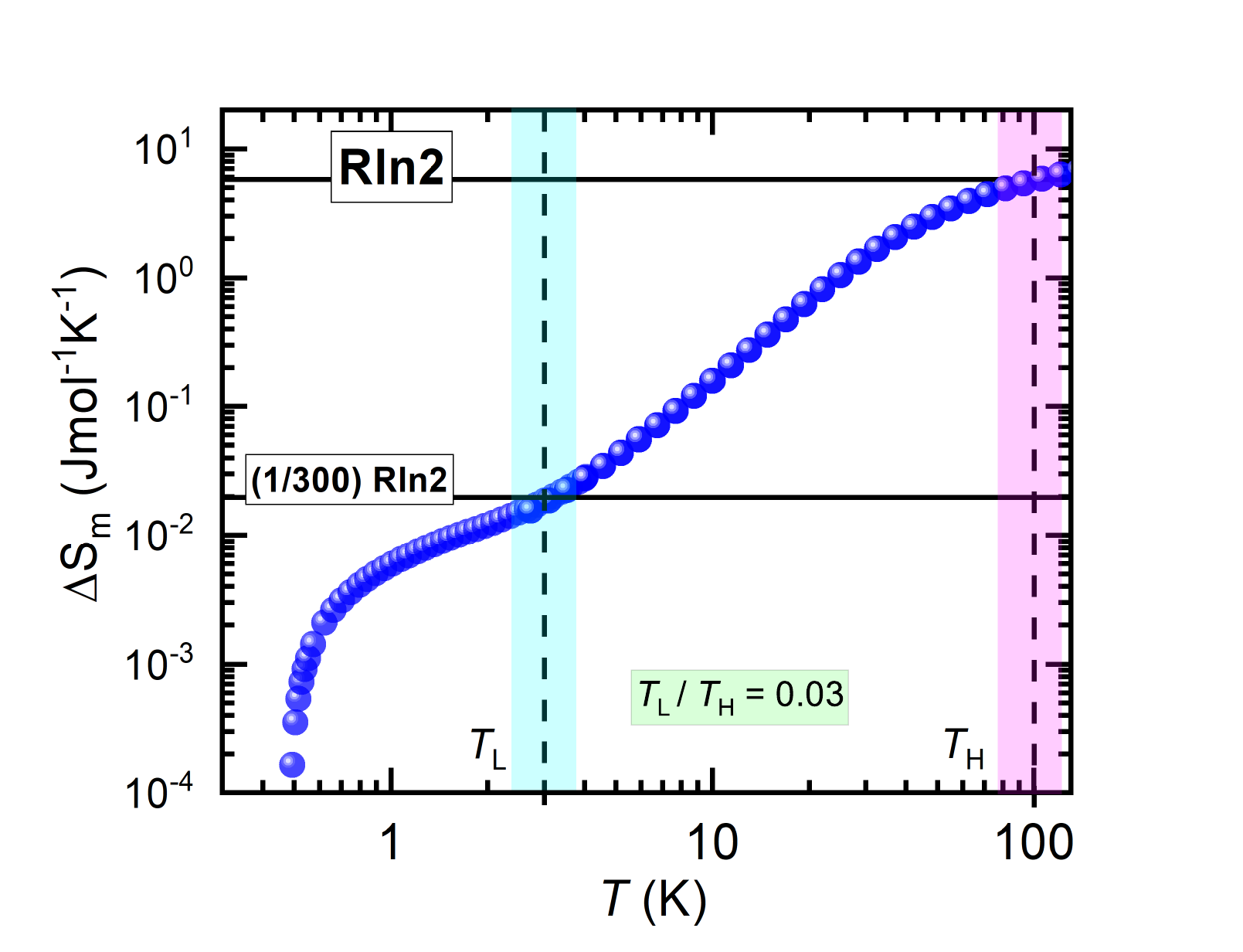}\caption{\label{Sm_HLRO_96h}{
\small{}Magnetic entropy change, $\Delta$S$_{m}$ as a function of temperature at 
zero field for \ch{H_{5.9}Li_{0.1}Ru_{2}O_{6}} (blue circles).  The entropy corresponding 
to a $J_{eff}=\frac{1}{2}$  magnetic system ($R \ln 2$) is  shown as a reference 
(the upper black horizontal line). Plateaus are seen at $T_{L}$  $\sim$ 3 K (vertical dotted line 
with light blue shade; from $Z_2$-flux excitations putatively) and 
$T_{H}$ (from itinerant Majorana excitations putatively) $\sim$ 100 K (vertical dotted line 
with pink shade). The ratio of $T_{L}$/$T_{H}$ is 0.03 consistent with isotropic Kitaev model.}
}
\end{figure}

From $C_{m}$, we  compute the magnetic 
entropy change $\Delta S_{m}$ 
for the system at zero field using $\Delta S_{m}(T) = \int^T_{T_{\text{lowest}}} 
\frac{C_{m}}{T'}dT'$ where $T_{\text{lowest}}$ is the lowest temperature measured
in our experiments (500 mK). \footnote{The magnitude of entropy change from low-$T$ to about 100 K ($\sim$ R$\ln$2) does not change much if one extrapolates the magnetic heat capacity to zero at 0 K.} Figure \ref{Sm_HLRO_96h} shows $\Delta$S$_{m}$ vs. $T$. By about 100 K, a magnetic
entropy value of $R \ln 2$ is recovered which is consistent with that for 
a $J_{\text{eff}}=1/2$ system. Further, we see a two-step entropy release for HLRO. 
For a Kitaev spin liquid, the two step entropy release 
corresponds to the freeing up of two types of quasi-particles -- $Z_{2}$ flux 
excitations at low-$T$ ($T_L$) and itinerant Majorana fermions at a higher-$T$ ($T_H$).
In the idealized case, the entropy  is released in two equal
halves of $R \ln 2 /2$ \cite{Nasu2015_2P} 
and the ratio $\frac{T_{L}}{T_{H}}$ $\sim$ 0.03\cite{Do2017,Nasu2015_2P,Motome2020_2P}.
We find that in HLRO the ratio $\frac{T_{L}}{T_{H}}$ $\sim$ 0.03 is 
consistent with the expected value, but the ratio  
for $\Delta S_{m}$ at the plateaus ($1/300$) does not match the expected value. 
We note that even for the field-induced KQSL in $\alpha$-RuCl$_{3}$ above 80 kOe, 
a similar two-step entropy release has been observed and proposed 
as an evidence of $Z_{2}$ flux and itinerant Majorana fermions,
but there are  devitations from expected values. 
The ratio of $\frac{T_{L}}{T_{H}}$ in $\alpha$-RuCl$_{3}$ is 0.22
and the appearance of $R \ln 2/2$ recovery at the lower-temperature plateau
due to flux excitations is also unclear~\cite{Do2017}. 
A small amount of vacancy and quasi-vacancy in HLRO might be affecting
the first plateau in the entropy release due to bound-flux sector excitations
as we have seen low temperature variation of $C_{m}/T \propto T^{-0.9}$ at zero field 
deviates from pure Kitaev limit ($C_{m}/T \propto T$)\cite{Perkins2021}. 
Alternately,
additional couplings such as Heisenberg and off-diagonal exchange terms 
has been shown as a possibility for the unequal entropy 
release~\cite{Yamaji_etal_PRB16, Catuneanu_etal_NPJ_2018, Li_etal_PRR20}.
Given the scalings observed in HLRO, the unequal entropy release 
may be ascribed to vacancy effects primarily.

\textit{Conclusion$-$} We propose HLRO to be a Ru$^{3+}$-containing  Kitaev QSL which shows no sign of long range ordering down to 84 mK and hence does not require a magnetic field to reveal Kitaev physics.
We observed a scaling behaviour of the magnetic specific heat as $C_{m}/T \propto (T/B)^\gamma$ which is consistent with defect/vacancy-induced effects predicted for a KQSL due to weak localization of Majorana fermions. An apparent divergence of $C_{m}/T$ with $T^{-0.9}$ at low-$T$ is also consistent with this picture.  A two-step entropy release is inferred from heat capacity data and $T_{L}/T_{H}\sim 0.03$, a key signature of KQSL. The deviation from R$\ln$2/2 in the entropy release at the first step expected for a pristine KQSL  is suggested to arise from defects/vacancies as well. Even though the presence of defects/vacancies is inevitable in a real material, here they
serve a positive purpose of probing the pristine state. This then opens up a new direction for future work where the effect of vacancy concentration on the scaling properties of Kitaev spin liquid can be investigated. The novel material system reported here can be especially suited for this. Thermal Hall effect and inelastic neutron scattering measurements in our system are also desirable for further validating our conclusions.

\section{acknowledgment}

We thank MOE STARS scheme for financial support.  
SP acknowledges funding support from SERB-DST, India via Grants No. MTR/2022/000386. We thank Eric Andrade, Ciar\'an Hickey, Edwin Kermarrec and Mohammad Monish for discussions. $\mu$SR experiments were performed at Paul Scherrer Institute (PSI), Switzerland with beam-time allocation ID: 20221240 on GPS, S$\mu$S facility. Dilution refrigerator $\mu$SR experiments were performed at ISIS, UK (https://doi.org/10.5286/ISIS.E.RB2410249).  We also thank MPI-CPFS, Dresden, Germany for financial support and access to measurement facilities. Thanks to IITB-central facilities for various measurements.

\bibliographystyle{plain}
\bibliography{citation}

\end{document}


\title{ Supplementary Material for \ch{(H,Li)_{6}Ru_{2}O_{6}}: a possible zero-field Ru$^{3+}$-based Kitaev Quantum Spin Liquid }

	\author{Sanjay Bachhar} 
	\email{sanjayphysics95@gmail.com}
	\affiliation{Department of Physics, Indian Institute of Technology Bombay, Powai, Mumbai 400076, India}

	\author{M. Baenitz}
	\affiliation{Max Planck Institute for Chemical Physics of Solids, 01187 Dresden, Germany}

	\author{Hubertus
		Luetkens}
	\affiliation{Laboratory for Muon Spin Spectroscopy, Paul Scherrer Institute, CH-5232 Villigen PSI, Switzerland}
	\author{John Wilkinson}
	\affiliation{ISIS Pulsed Neutron and Muon Source, STFC Rutherford Appleton Laboratory,
		Harwell Campus, Didcot, Oxfordshire OX110QX, United Kingdom}
	\author{Sumiran Pujari}
	\affiliation{Department of Physics, Indian Institute of Technology Bombay, Powai, Mumbai 400076, India}

	\author{A.V. Mahajan}
	\email{mahajan@phy.iitb.ac.in}
	\affiliation{Department of Physics, Indian Institute of Technology Bombay, Powai, Mumbai 400076, India}

\date{\today}
\maketitle

 Polycrystalline \ch{(H,Li)_{6}Ru2O6} are prepared through solid-state reaction followed by Hydrothermal method. X-ray diffraction evidences a single phase pure compound. The Ru-valency was determined by X-ray photoelectron spectroscopy (XPS). A systematic NMR spectral intensity calculation suggested the Li-content in the compound and H-content was estimated through charge balance. Then, we have measured magnetization, heat capacity and local probe (Nuclear Magnetic Resonance (NMR) and Muon Spin Relaxation ($\mu$SR)) to understand underlying physics of \ch{(H,Li)_{6}Ru2O6}. Herein, we report supplementary information of \ch{(H,Li)_{6}Ru_{2}O_{6}}:  a zero-field Ru$^{3+}$-based Kitaev Quantum Spin Liquid.

\section{Sample Preparation}

The preparation of \ch{(H,Li)_{6}Ru2O6} samples, involves a two steps process, which was carried out at IIT Bombay facility. \ch{Li2RuO3}, precursor is prepared using the standard solid-state reaction method. Then, H-intercalation was carried out in \ch{Li2RuO3} using a hydrothermal method (similar method was used in preparation of \ch{H3LiIr2O6}, a KQSL \cite{Bette2017}) that applies both pressure and temperature to the compound. This process do replacement of Li with H, forming a new compound. Depending on intercalation time, Li-quantity will vary (for details see \ref{D-NMR-HLRO96h} ) in resultant compound. The compound with 96 hours intercalation time, was carried out, the reduction of Li in the final sample was confirmed through a flame test of the by-product (which contained the extracted Li from \ch{Li2RuO3}), and the structure was analyzed using X-ray diffraction. The valency of Ruthenium was determined using XPS, which showed that it was in the Ru$^{3+}$ state. This led to the discovery of a reproducible 2D layered compound, \ch{(H,Li)_{6}Ru2O6}, which could serve as an alternative to $\alpha$-\ch{RuCl3}. To determine the amount of Li in \ch{(H,Li)_{6}Ru2O6}, we used the $^{7}$Li-NMR Spectral intensity method, as described in section \ref{D-NMR-HLRO96h}. Our investigation has shown that the relationship between intercalation time and Li content for Ir and Ru-based compounds is different. For the compound with 96 hours intercalation time, we found that the Li content was approximately 2.5{\%} relative to \ch{Li2RuO3}, leading to the chemical formula \ch{H_{5.9}Li_{0.1}Ru2O6} (HLRO) under charge balance. 

\section{X-ray diffraction and Rietveld Refinement}

To check the phase purity of HLRO, we have measured the powder x-ray diffraction (XRD) of polycrystalline sample at IIT Bombay central facilities. The XRD data were collected at room temperature with Cu-K$_{\alpha}$ radiation ($\lambda$=1.54182 \AA) over the angular range 10{\textdegree} $\leq$ 2$\theta$ $\leq$ 90{\textdegree} with a 0.013{\textdegree} step size. 

\begin{figure}[!h]
	\centering
	\includegraphics[width=1.0\linewidth]{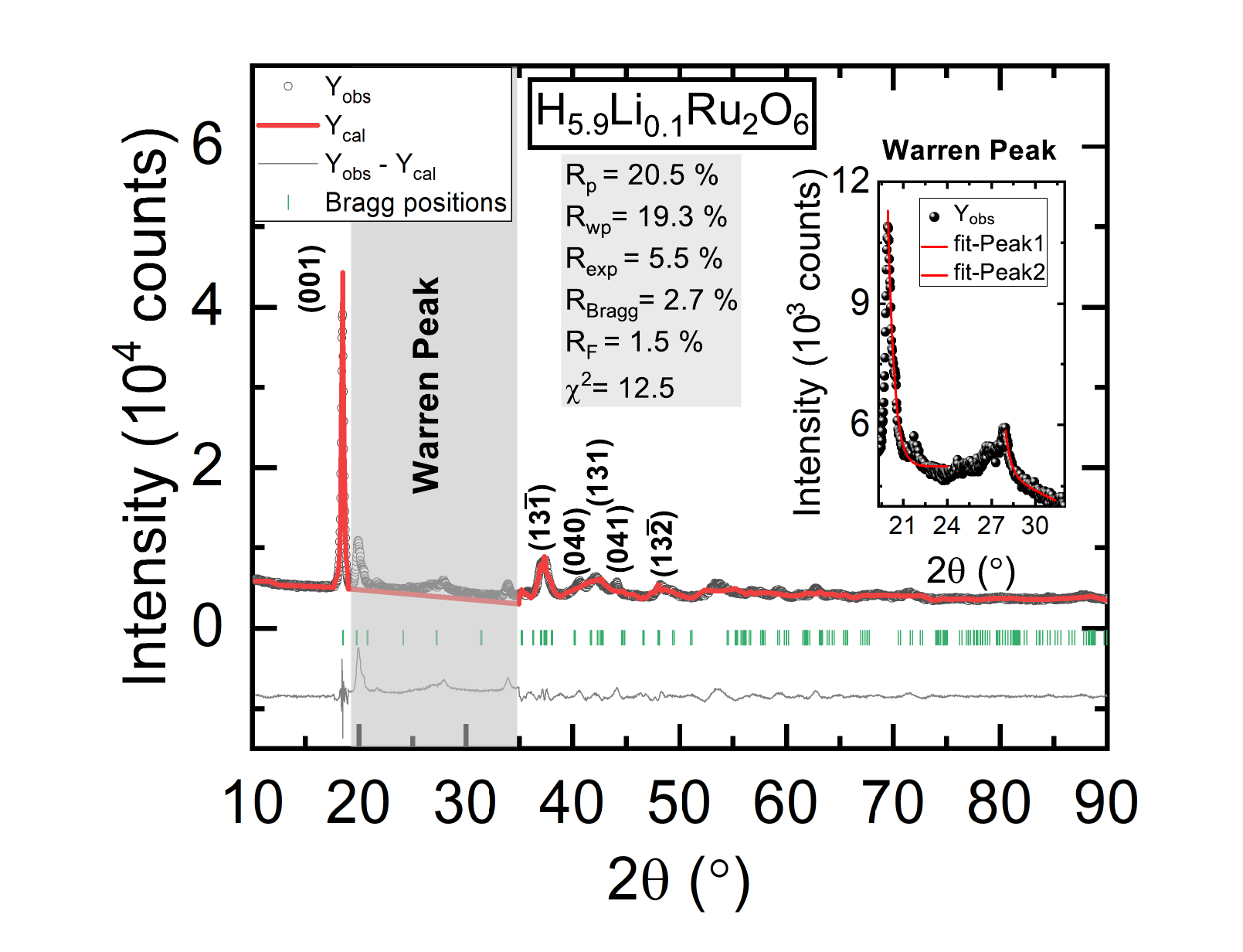}
	\caption{The single phase Rietveld refinement of XRD pattern of \ch{H_{5.9}Li_{0.1}Ru2O6}. The open black circles are the observed data, the red line is theoretically calculated one, the grey line is the difference between the observed and the calculated data and the green vertical marks are Bragg positions with corresponding Miller indices. Inset shows the Warren peak.}
	\label{XRDrefinement_HLRO_96h}
\end{figure}

Figure \ref{XRDrefinement_HLRO_96h} shows the single phase Rietveld refinement of HLRO. After the refinement, we found that the prepared HLRO crystallizes in the monoclinic C2/m space group with lattice parameter $a=5.07$ \AA, $b=8.97$ \AA, $c=5.01$ \AA, $\alpha=\gamma=90${\textdegree} and $\beta=106.8${\textdegree}. The goodness of the Rietveld refinement is defined by the following parameters; $\chi^{2}=12.5$, $R_{p}=20.5{\%}$, $R_{wp}=19.3{\%}$, $R_{exp}=5.5{\%}$. Table \ref{table1} summarizes the unit cell paramters and quality factors for the Rietveld refinement of HLRO. A Warren peak is observed in the range 19{\textdegree} to 35{\textdegree}, shown in inset of Figure \ref{XRDrefinement_HLRO_96h} with a fit \cite{Bahrami_T_2019} as given below,

\begin{equation}
	I_w(2\theta)= A e^{-g(2\theta)^2}+ B/(C+(2\theta)^2),	
\end{equation}

where A, B and C are constants, $ g $ is the exponent of the Gaussian term and it measures the
percentage of the stacking faults known as the $ g $-factor:
\begin{equation}
	g=\delta^2/d^2, 	\delta^2 = <d^2>-<d>^2
\end{equation}
where $ d $ is the interlayer spacing. A good fit with exponent, $ g=0.06(6) $ in the inset of Figure \ref{XRDrefinement_HLRO_96h}  corresponds to at least 6\% volume fraction of
stacking disorder. After excluding the Warren peak, Rietveld refinement is performed with faultless model (stacking faults are ignored) to extract different quality factors, atomic coordinates, site occupancies, and the isotropic Debye-Waller factors (B$_{iso}$ = 8$\pi^2$U$_{iso}$) of  HLRO,which are tabulated in Table \ref{table1} and Table \ref{table2} respectively.

\begin{table}[h!]
	\caption{Unit cell parameters and quality factors are reported for the Rietveld refinement of \ch{H_{5.9}Li_{0.1}Ru2O6} at room temperature.}
	\centering
	\scalebox{0.7}{
		\begin{tabular}{|l c| c c|}
			\hline
			Unit Cell Parameters for C2/m& & Quality Factors &\\
			\hline
			a(\AA) & 5.07(3) &  &  \\
			\hline
			b(\AA)& 8.97(4)& R$_{Bragg}$ (\%)&2.7 \\
			\hline
			c(\AA)& 5.01(2) & R$_F$ (\%) &1.5 \\
			\hline
			$\alpha$=$\gamma$  (\textdegree)&90 & R$_{exp}$ (\%)&5.5\\
			\hline
			$\beta$ (\textdegree)& 106.8(8) & R$_p$ (\%)&20.5\\
			\hline
			Z  & 2 & R$_{wp}$ (\%) & 19.3\\
			\hline
			V (\AA$^3$) & 218.3(4) & $\chi^2$ & 12.5\\
			\hline	
		\end{tabular}
	}
	\label{table1}
\end{table}

\begin{table}[h!]
	\caption{Atomic coordinates, Normalized site occupancies, and the
		isotropic Debye-Waller factors (B$_{iso}$ = 8$\pi^2$U$_{iso}$)  are reported
		for the Rietveld refinement of \ch{H_{5.9}Li_{0.1}Ru2O6}.}
	\centering
	\scalebox{0.7}{
		\begin{tabular}{|l| c| c| c| c| c| c|c|}
			\hline
			Atom & Wyckoff position & Site &x&y&z&Norm. Site Occ.&B$_{iso}$(\AA$^2$)\\
			\hline
			Ru(1)&4g& 2&0&0.333&0&2&0.1\\
			\hline
			Li(1)&2a& 2/m&0&0&0&0.1&0.1\\
			\hline
			O(1)&4i&m&0.417&0&0.22&2&1.7\\
			\hline
			O(2)&8j&1&0.404&0.323&0.229&4&1.7\\
			\hline
			H(1)&4h&2&0&0.161&1/2&2&0.1\\
			\hline
			H(2)&2d&2/m&1/2&0&1/2&3&0.1\\
			\hline
			H(3)&2a&2/m&0&0&0&0.9&0.1\\
			
			\hline	
		\end{tabular}
	}
	\label{table2}
\end{table}

\section{X-ray Photoelectron Spectroscopy (XPS)}

X-ray photoelectron spectroscopy (XPS) is a surface-sensitive technique that is based on the photoelectric effect. It is used to determine the elemental composition and chemical state of a material, as well as the electronic structure and density of the electronic states in the material. XPS is a powerful technique because it not only shows which elements are present, but also how they are bonded to other elements. The chemical states are inferred from the binding energy and number of ejected electrons. X-ray photoelectron spectroscopy (XPS) is also one of the most popular method to investigate valency of elements present in a compound. In order to know valency of Ru in HLRO, XPS spectra were measured at Central Surface Analytical Facility (ESCA lab) at IIT Bombay.  The C1s reference position was taken as 284.7 eV for all the materials. After modeling and fitting the spectra in ESCape software, the data was exported to an Excel sheet and plotted in Origin software. RuO$_{2}$ (precursor of HLRO) which is a good example of Ru$^{4+}$ compound, was measured to use as a reference to investigate XPS spectra of HLRO.

\begin{figure}[!h]
	\centering
	\includegraphics[width=1.0\linewidth]{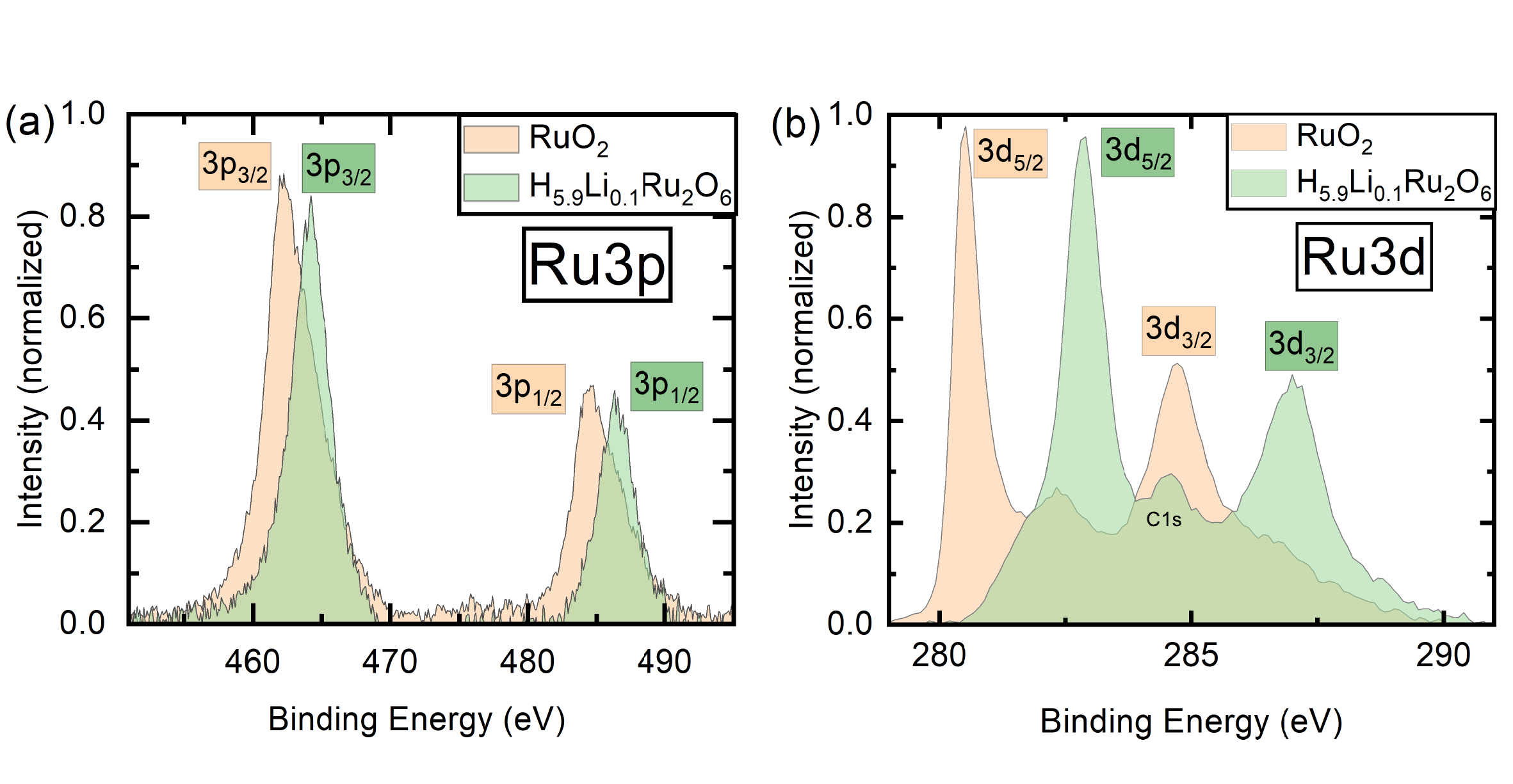}
	\caption{(a) Ru3p XPS spectra of \ch{H_{5.9}Li_{0.1}Ru2O6} with reference to RuO$_2$. (b) Ru3d XPS spectra of \ch{H_{5.9}Li_{0.1}Ru2O6} with reference to RuO$_2$.}
	\label{Ru3dHLRO}
\end{figure}

\begin{table}[h!]
	\caption{ Binding energies for \ch{H_{5.9}Li_{0.1}Ru2O6}, RuO$_{2}$ and RuCl$_{3}$.}
	\centering
	\scalebox{0.7}{
		\begin{tabular}{|c|c|c|c|c|}
			\hline
			Material & Peak & Binding Energy (eV) & Oxidation State & Comments\\
			\hline
			HLRO & 3d$_{5/2}$ &282.9&& Present Study.\\
			\cline{2-3}
			&3d$_{3/2}$&287&3+&\\
			\cline{2-3}	
			&3p$_{3/2}$&464.2&&\\
			\cline{2-3}	
			&3p$_{1/2}$&486.3&&\\
			\hline
			RuCl$_{3}$ & 3d$_{5/2}$ &282.8&& Taken  \\
			\cline{2-3}
			&3d$_{3/2}$&287&3+& from reference \cite{Morgan2015}.\\
			\cline{2-3}	
			&3p$_{3/2}$&464.1&&\\
			\hline
			RuO$_{2}$ & 3d$_{5/2}$ &280.5&& Present Study and \\
			\cline{2-3}
			&3d$_{3/2}$&284.7&4+& also its consistent \\
			\cline{2-3}	
			&3p$_{3/2}$&462.2&& with reference \cite{Morgan2015}.\\
			\cline{2-3}	
			&3p$_{1/2}$&484.7&&\\
			\hline
		\end{tabular}
	}
	\label{XPS_Table_3}
	
\end{table}

Figure \ref{Ru3dHLRO}(a) shows $Ru3p$ XPS spectra of  HLRO with reference to RuO$_{2}$. There is clear right shift in $3p_{3/2}$ and $3p_{1/2}$.  Figure \ref{Ru3dHLRO}(b) shows $Ru3d$ XPS spectra of  HLRO with reference to RuO$_{2}$. There is also clear right shift in $3d_{5/2}$ and $3d_{3/2}$. Shift in both Ru3p and Ru3d spectra indicates valency of Ru in  HLRO is not in 4+ state. We analysed XPS spectra by using ESCape software to extract Binding energies of different peaks and it is shown in table \ref{XPS_Table_3}. Binding energies of Ru$3d_{5/2}$, Ru$3d_{3/2}$ and Ru$3p_{3/2}$ in HLRO are consistent with RuCl$_{3}$ where Ru is in 3+ state as reported in literature \cite{Morgan2015}. That indicates Ru in  HLRO is in 3+ state.

\section{Crystal Structure}

\begin{figure}[!h]
	\centering
	\includegraphics[width=1\linewidth]{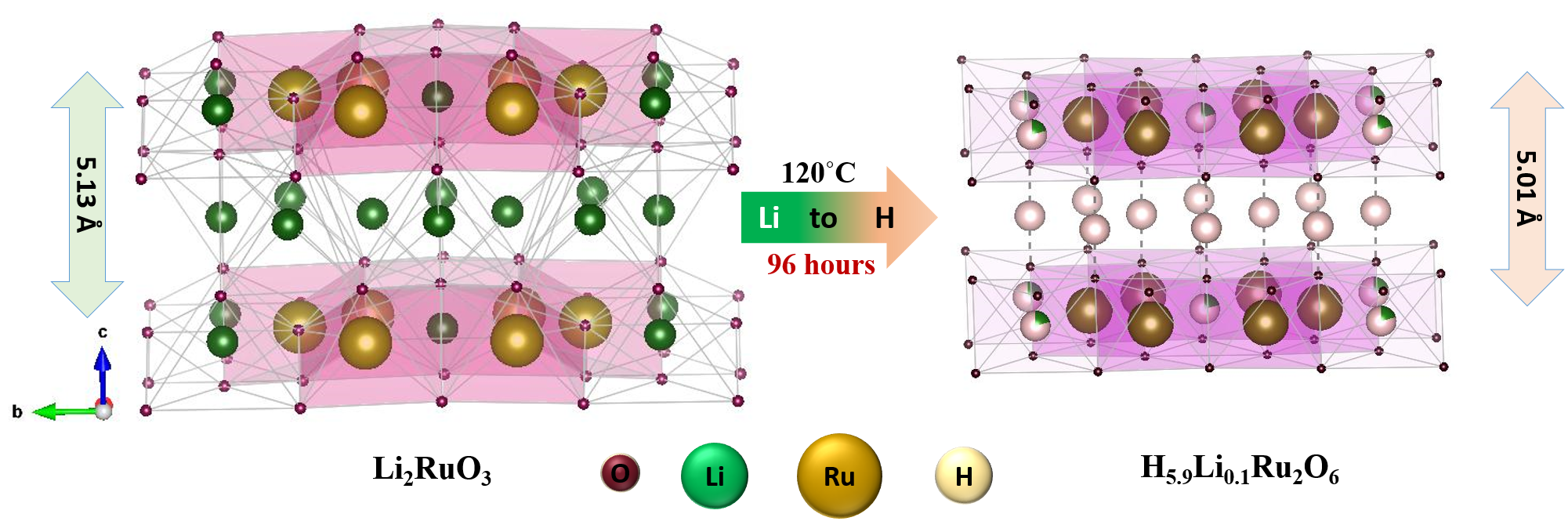}
	\caption{The crystal structure of Li$_{2}$RuO$_{3}$ (left) that
		can be transformed into H$_{5.9}$Li$_{0.1}$Ru$_{2}$O$_{6}$  (right) by 96 hours H-intercalation (Hydrothermal Method).}
	\label{Crystal-structure}
\end{figure}

Figure \ref{Crystal-structure} depicts the crystal structures of \ch{H_{5.9}Li_{0.1}Ru{2}O{6}} (space group $C2/m$) and its precursor, \ch{Li2RuO3} (space group P21/m). The precursor, \ch{Li2RuO3}, undergoes strong dimerization at temperatures below $\sim$ 540 K, resulting in the formation of molecular orbitals (MOs)\cite{Yoko2007}. In contrast, intercalation has been found to suppress Ru-Ru dimerization in \ch{Ag3LiRu2O6} \cite{Kimber2010,Takagi2022,RKumar2019}. The H-intercalation process for 96 hours at 120{\textdegree}C led to the replacement of Li atoms by H atoms from both the inter-layer positions and the in-plane positions (partially). The contraction of the c-parameter of the lattice (from 5.13 {\AA} to 5.01 {\AA}) due to the lighter H atoms replacing the inter-layered Li atoms is illustrated in Figure \ref{Crystal-structure}. The structural analysis reveals a perfect honeycomb network with a $Ru-Ru$ bond distance of $\sim$ 2.96 {\AA} and a $d-p-d$ bond angle of $\sim$ 95{\textdegree}. The presence of Ru$^{3+}$ (as indicated by XPS analysis) at the vertices of this perfect honeycomb network makes it a Kitaev honeycomb system, serving as an alternative to $\alpha$-\ch{RuCl3}.

\section{Magnetization}

The dc magnetization $M(T)$ as a function of temperature ($T$, range 2-300 K) was measured on a cold pressed pellet of HLRO in zero field cooled (ZFC) at 10 kOe field, also in zero field cooled (ZFC)-field cooled (FC) mode at 50 Oe and $M(H)$ vs $H$ (magnetic field) at 20 K using a Quantum Design Magnetic Property Measurement System (MPMS). The main features of our observations from the measurement are described below.

\begin{figure}[!h]
	\centering
	\includegraphics[width=1.0\linewidth]{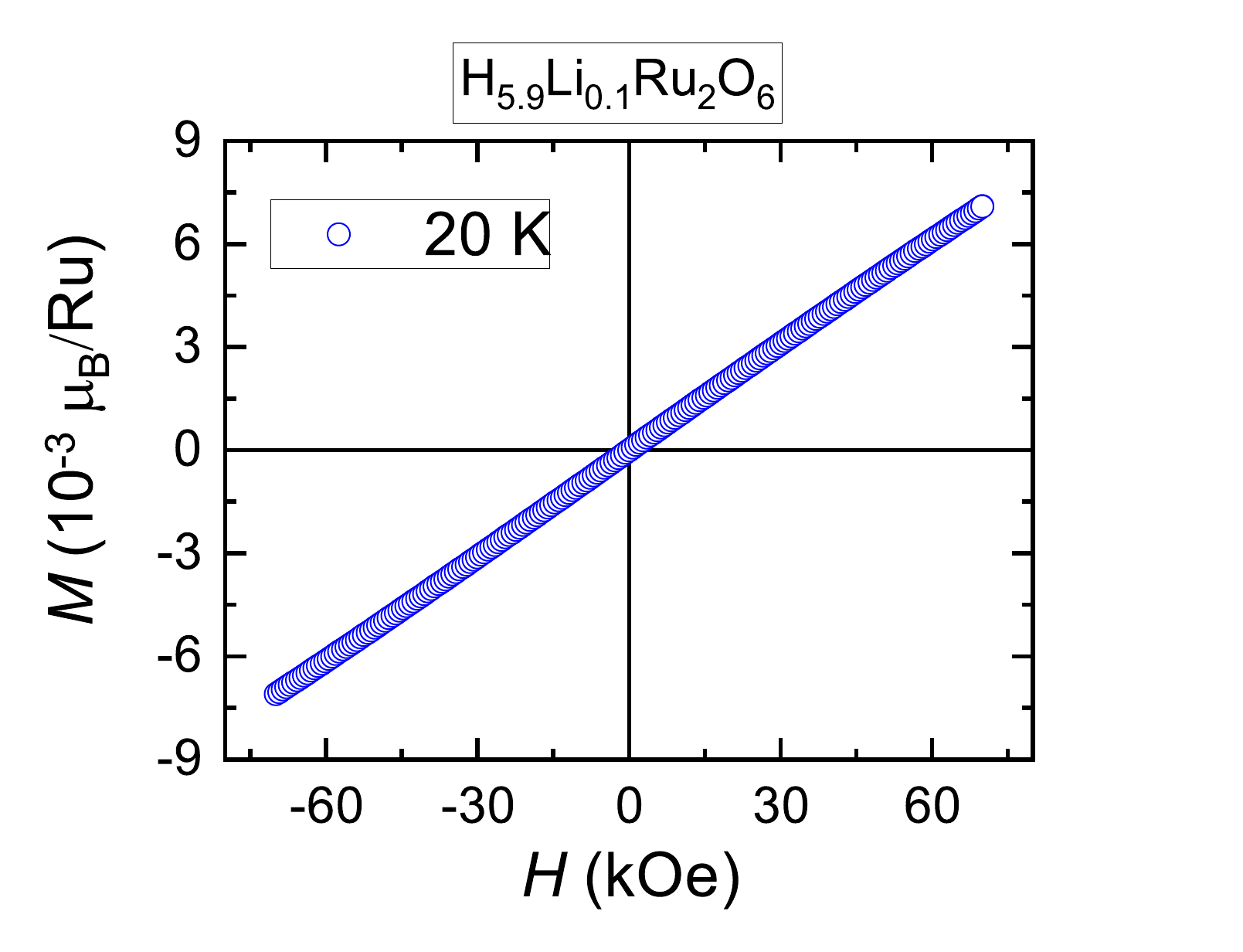}
	\caption{$M$ vs $H$ at 20 K in the field range ($-$70 kOe to 70 kOe).}
	\label{MH_HLRO96h}
\end{figure}

\begin{figure}[!h]
	\centering
	\includegraphics[width=1.0\linewidth]{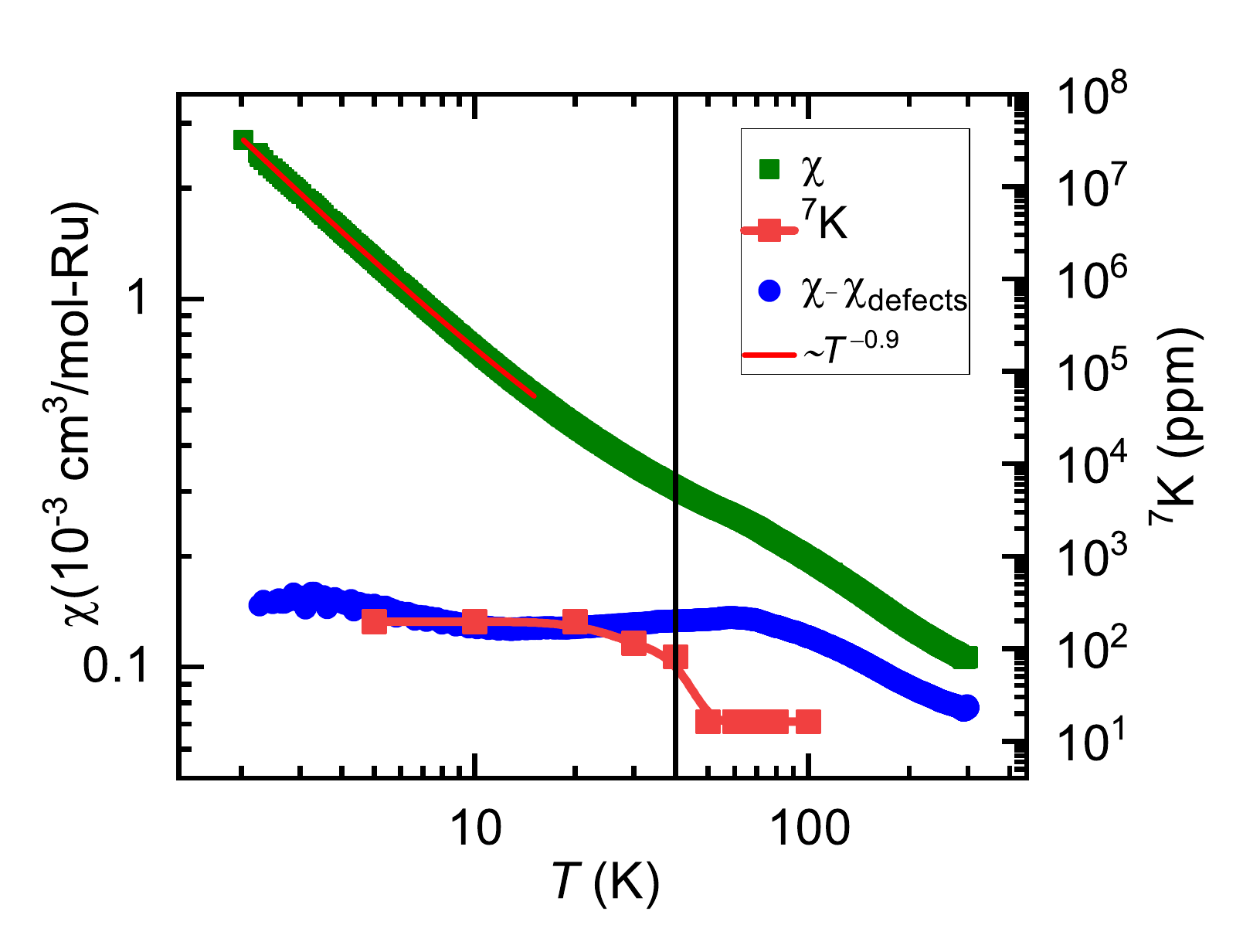}
	\caption{A log-log plot for the  magnetic susceptibility $\chi$ is shown (left y-axis, green squares) together with the NMR shift $^{7}$K (right y-axis, red squares). The low-$T$ power-law fit ($\sim T^{-0.9}$) has been shown with a red solid line. After subtraction of a power-law fit for the low-$T$ part, we plotted  $\chi-\chi_{defects}$. This then follows the $T$-variation of the NMR shift $^{7}$K.}
	\label{KChi_HLRO96h}
\end{figure}

\begin{figure}[!h]
	\centering
	\includegraphics[width=1.0\linewidth]{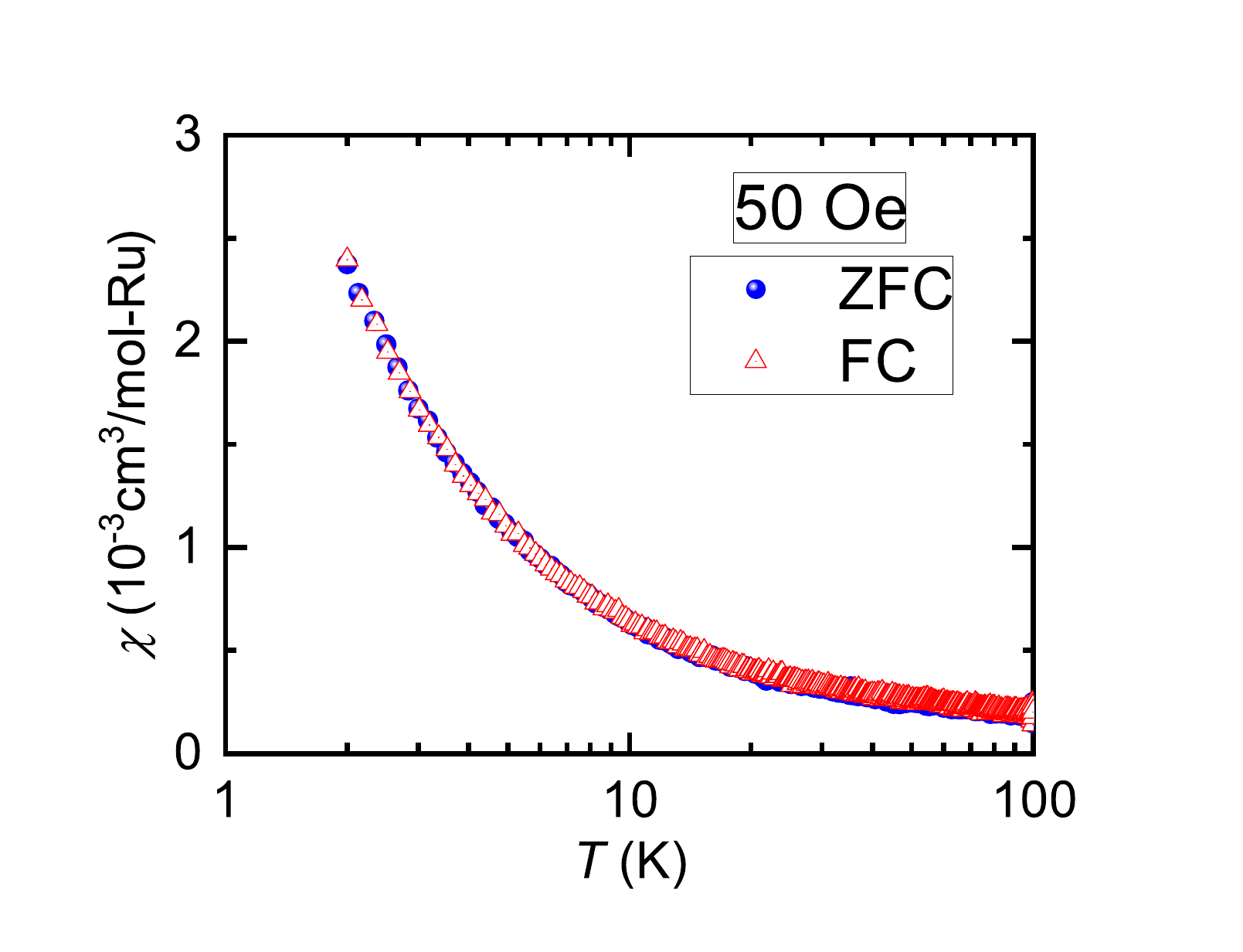}
	\caption{No bifurcation in the the semi-log plot of $\chi(T)$ vs. $T$ between zero field cooled (ZFC) (blue sphere) and field cooled (FC) (red triangle) mode at an applied field $H=50$ Oe in the temperature range 100 K to 2 K.}
	\label{HLRO_MT_50Oe}
\end{figure}

Figure \ref{MH_HLRO96h} shows $M$ vs $H$ at 20 K in the field range (-70 kOe to 70 kOe). We observe that it is a perfect linear isotherm ($M(H)= \chi H$, where $\chi$ is susceptibility) passing through origin (0,0) of M vs H at 20 K plot. That indicates there is no ferromagnetic impurity present in system. Then, we calculated $\chi$ by using $\chi=M/H$. The bulk magnetic susceptibility, $\chi$ shows a cross over around 40 K which has also been observed in other techniques such as specific heat, NMR and $\mu$SR, as can be found in the respective sections later. No sharp anomaly down to 2 K has been seen in $\chi$ vs $T$ (300 K to 2 K) (Figure \ref{KChi_HLRO96h}). This feature rules out the presence of any kind of either long-range ordering (LRO) or frozen magnetism within the system. Curie-Weiss fitting ($\chi= \chi_{0}$ + $\frac{C}{T-\theta_{CW}}$) in the temperature range 100-300 K (black solid line on red square symbol of 1/($\chi-\chi_{0}$) vs. $T$ plot in main paper-Fig.2) reveals $\chi_{0}= 3.69 \times 10^{-5}$ cm$^{3}$/mol-Ru, Curie-Weiss temperature $\sim -44$ K and effective moment $\sim$ 0.43 $\mu_B$ which is 4 times smaller than the effective moment corresponding to a pure $J_{eff}=1/2$ system. Figure \ref{KChi_HLRO96h} shows a power-law fit with $\chi \sim T^{-\eta}$ ($\eta=0.9$). We ascribe this to the presence of defects in the system\cite{Andrade2022,Perkins2021}. It is rather challenging to distinguish such a intrinsic defect contribution from a low-$T$ Curie contribution (due to extrinsic defects) since the exponent ($\eta=0.9$) is close to 1. However, magnetic heat capacity behavior at low-$T$ is consistent with the predictions of the effect of vacancy in KQSL\cite{Andrade2022,Perkins2021}. Hence, we suggest that the defects are of intrinsic type and originate from vacancies and quasi-vacancies. After subtraction of power-law fit, we observed a nearly temperature independent-$\chi$ like the Pauli susceptibility. The NMR shift which is proportional to the intrinsic spin susceptibility   is also temperature-independent at low-$T$. Observed $^{7}$Li/$^{1}$H NMR lineshape largely comes from defect-free pristine region and has marginal contribution from defects which gives the power-law in the susceptibility. No bifurcation in plot of $\chi(T)$ vs. $T$ between ZFC-FC mode at an applied field $H=50$ Oe (Figure \ref{HLRO_MT_50Oe}) in the temperature range 100 K to 2 K rules out possibility of static/glassy moment present in the system.

\section{Heat Capacity}

To get more information about low-energy excitations, we measured the heat capacity of the sample at constant pressure $C_{p}(T)$ in the temperature range (500 mK-300 K) at 0 kOe and in the temperature range (400 mK-40 K) at 10-60 kOe for HLRO and non-magnetic analog \ch{Li3LiSn2O6} in the temperature range 1.8-300 K.

In general total specific heat, $C_{p}$ of the system can be expressed as:

\begin{equation}
	C_{p}(T,H)= C_{lattice}(T) + C_{Sch}(T,H) + C_{m}(T,H)
\end{equation}

The lattice specific heat is denoted as $C_{lattice}$, while $C_{Sch}$ represents the Schottky contribution caused by independent paramagnetic spins present in the system. The remaining magnetic contribution to the specific heat is referred to as $C_{m}$. Our main objective is to determine the "intrinsic" magnetic contribution $C_{m}$, which is independent of $C_{lattice}$ and $C_{Sch}$. To obtain the intrinsic $C_{m}$, we must first calculate $C_{lattice}$ and $C_{Sch}$, and then deduct them from the total specific heat $C_{p}$. There are a few ways to calculate $C_{lattice}$. A prominent way is to measure heat capacity of a suitable non-magnetic analog and another way is Debye-Einstein fitting\cite{SKundu2020_YCTO}.

\begin{figure}[!h]
	\centering
	\includegraphics[width=1.0\linewidth]{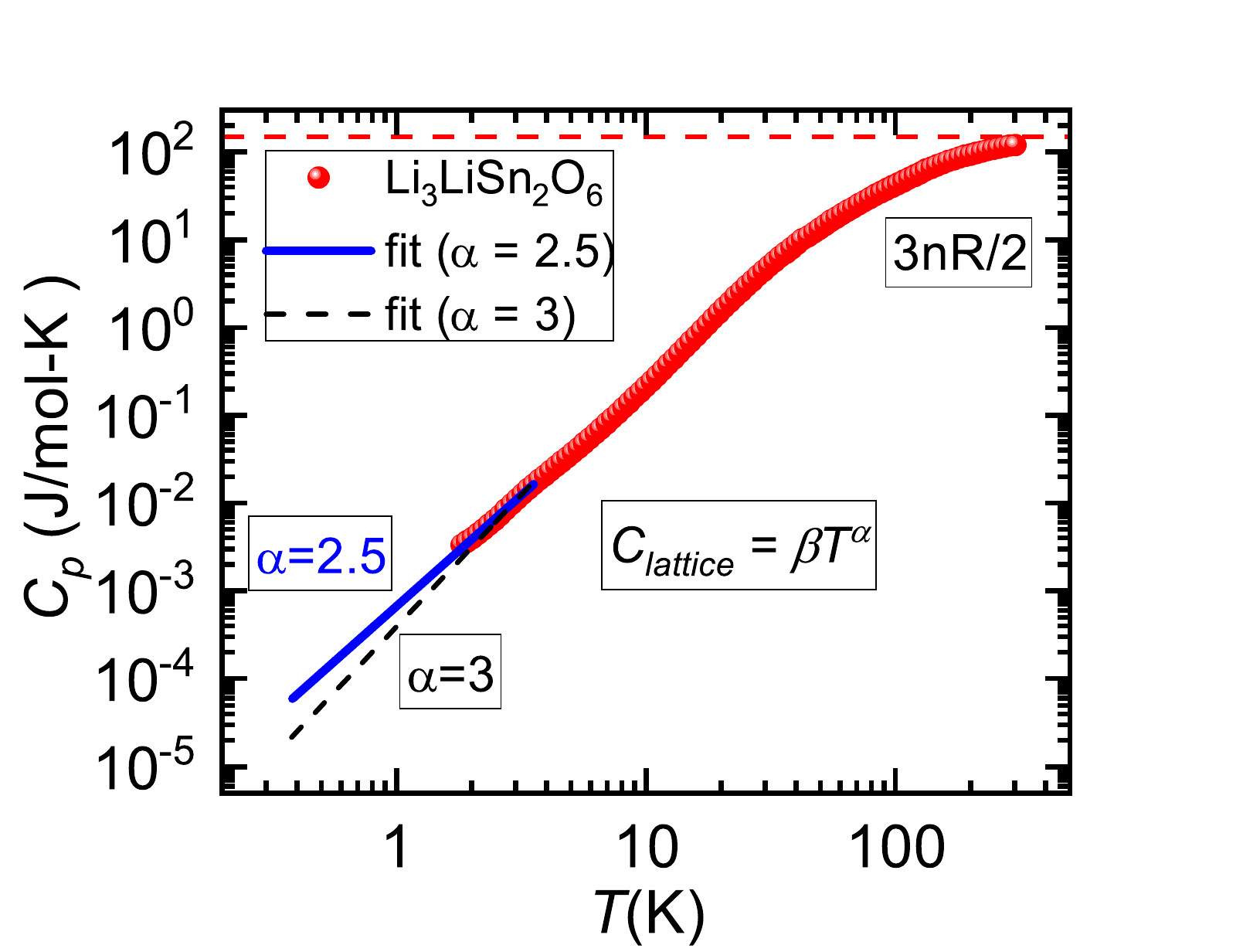}
	\caption{Specific heat, $C_{p}$ per Sn as a function of temperature for Li$_{3}$LiSn$_{2}$O$_{6}$ (red sphere). blue solid line and black dot line are fitted curve with $\alpha=2.5$ and $\alpha=3$ respectively.}
	\label{Cp_LSO}
\end{figure}

\begin{figure}[!h]
	\centering
	\includegraphics[width=1.0\linewidth]{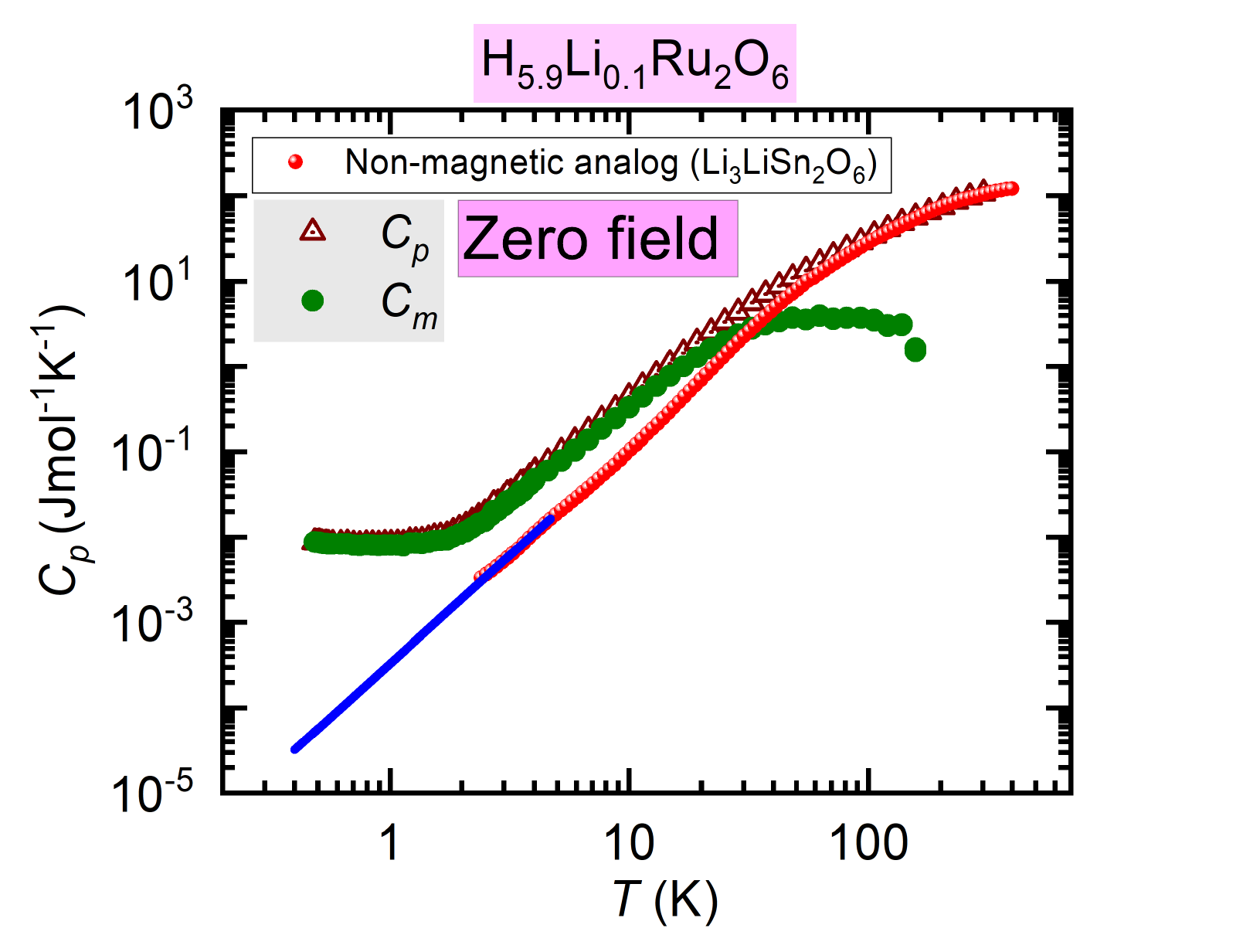}
	\caption{Specific heat, $C_{p}$ per Ru as a function of temperature for \ch{H_{5.9}Li_{0.1}Ru_{2}O_{6}} (wine triangle). The non-magnetic analog is shown by red sphere followed by blue line (extrapolated part). Magnetic heat capacity, $C_{m}$ is also shown in the same plot by olive circle.}
	\label{Cp_HLRO96h}
\end{figure}

\begin{figure}[!h]
	\centering
	\includegraphics[width=1.0\linewidth]{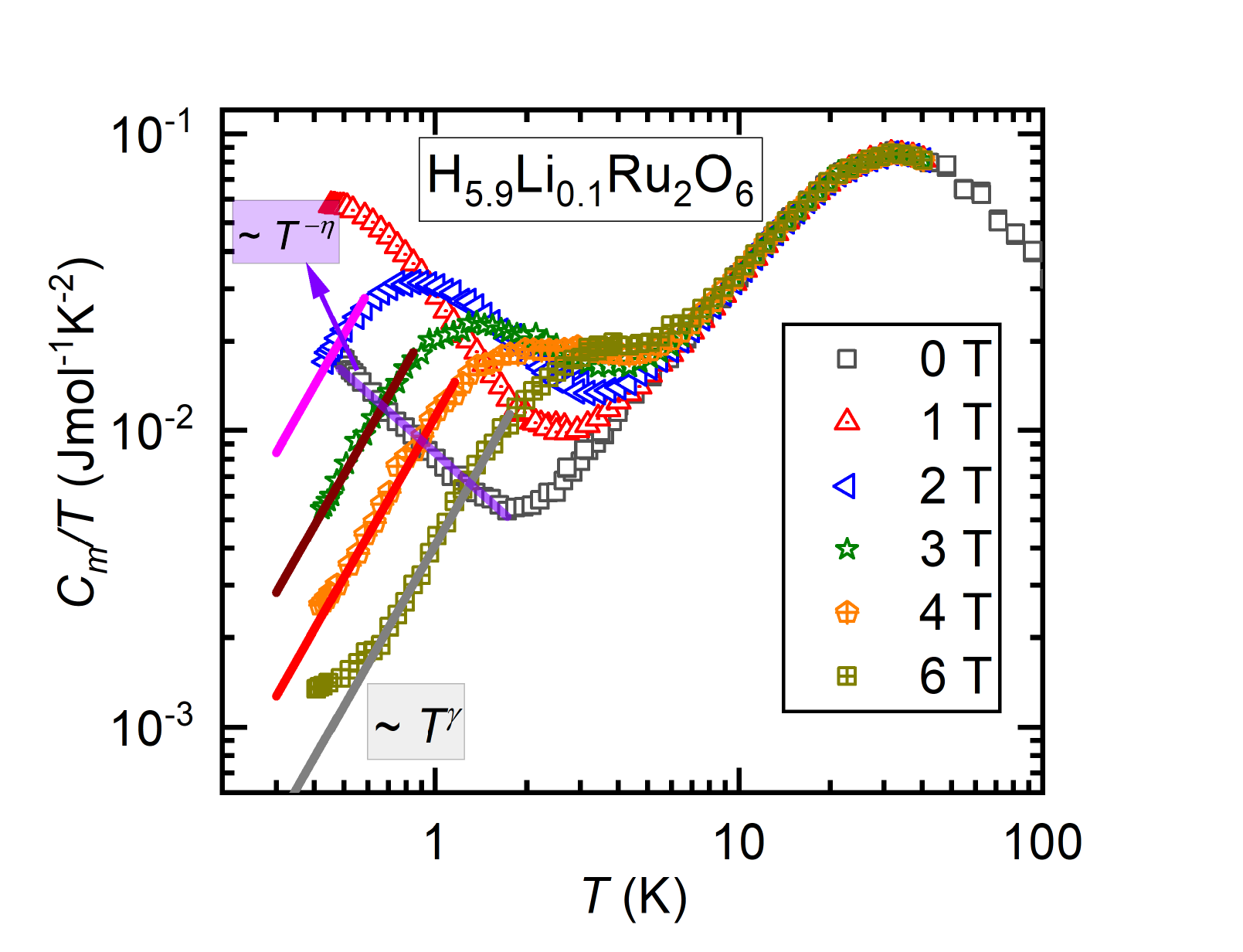}
	\caption{$C_{m}/T$ versus $T$ for various fields: 0-6 Tesla where $C_{m}$ is the magnetic heat capacity. Below about 2 K, it follows $T^{-\eta}$ with $\eta=0.9$ in zero field. With application of a field, it follows a  power-law ($\sim T^\gamma$ with $\gamma=1.8$) at low-$T$.}
	\label{Cm_HLRO96h}
\end{figure}

\begin{figure}[!h]
	\centering
	\includegraphics[width=1.0\linewidth]{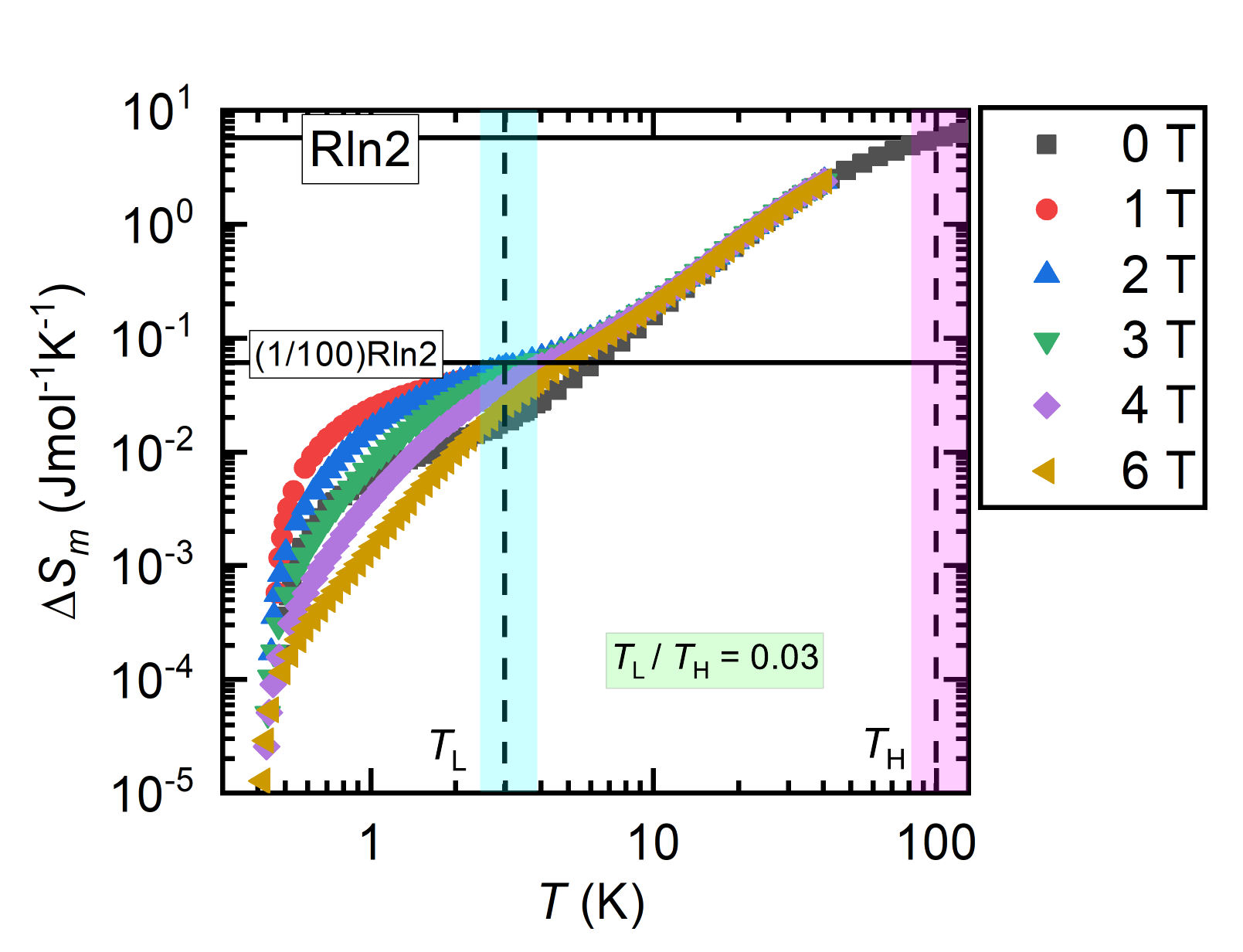}
	\caption{Magnetic entropy change, $\Delta$S$_{m}$ as a function of temperature at various fields: 0-6 T for \ch{H_{5.9}Li_{0.1}Ru_{2}O_{6}}.  The entropy corresponding to a $J_{eff}$=$\frac{1}{2}$  magnetic system ($Rln2$) is  shown as a reference (the upper black horizontal line).}
	\label{Sm_fields}
\end{figure}

We have a suitable non-magnetic analog \ch{Li3LiSn2O6} (see Figure \ref{Cp_LSO}) as a heat capacity reference of HLRO. We have measured the heat capacity of \ch{Li3LiSn2O6} from 300 K to 1.8 K. At high temperature, value of $C_{p}$ should be 3nR (n is number of atoms per fomula unit, R is gas constant) according to Dulong-Petit law\cite{Kittel2005}. We plotted $C_{p}$ per Sn in Figure \ref{Cp_LSO}, so we expected 3nR/2 at 300 K. We observe $C_{p}$ of \ch{Li3LiSn2O6} at 300 K matching with 3nR/2 reference line (pink dash line).  At low-$T$, lattice heat capacity should follow power law ($C_{lattice}\sim \beta T^{3}$) theoretically (black dotted line is shown in Figure \ref{Cp_LSO}). We fitted data to $C_{lattice} \sim \beta T^{\alpha}$ in the temperature range 1.8-3.6 K and it yields $\beta=0.0007$ and $\alpha=2.5$. Then, we extrapolated the data from 1.8 K to 400 mK for $C_{m}$ calculation. Note that at 500 mK, $C_{p}$ of \ch{Li3LiSn2O6} value $\sim$ 10$^{-4}$ which is 1/100 th of measured $C_{p}$ of HLRO (that is $\sim$ 10$^{-2}$). So, small error ($\Delta$ $\alpha$ $=$ 0.5) in $C_{lattice}$ at low-$T$ does not affect $C_{m}$ much.

Figure \ref{Cp_HLRO96h} shows specific heat, $C_{p}$ as a function of temperature for \ch{H_{5.9}Li_{0.1}Ru2O6} (wine triangle), non-magnetic analog \ch{Li3LiSn2O6} (red sphere). Magnetic heat capacity, $C_{m}$ (olive circle) for \ch{H_{5.9}Li_{0.1}Ru2O6} is shown at the same plot in log-log scaling to clearly show low-T features. The specific heat of the magnetic and non-magnetic analogs is expected to match at high temperatures (typically 100-300 K) with scaling of the Debye-temperature, which depends on the molar mass of the compound. In zero field, $C_{m}$ is the difference between as-measured $C_{p}$ and $C_{lattice}$ (that is $C_{p}$ of nonmagnetic analog) since $C_{sch} = 0$ at zero field. We calculated $C_{m}$ and it is shown in the Figure \ref{Cp_HLRO96h} with as-measured $C_{p}$ of \ch{H_{5.9}Li_{0.1}Ru2O6} and $C_{p}$ of \ch{Li3LiSn2O6}. No anomaly down to 500 mK indicates absence of long range magnetic ordering. In \ch{H3LiIr2O6}, there is a power law increase of $C_{p}/T \propto T^{-1/2}$ and does not approach zero at $T=0$, indicating presence of highly degenerate low-lying excitations around $E = 0$. The $C_{p}$ and $C_{m}$ at low-$T$ are nearly the same because lattice contribution is negligible (below 1 K, it is $\sim$ 10$^{-3}$ to 10$^{-5}$ Jmol$^{-1}$K$^{-1}$). In \ch{H_{5.9}Li_{0.1}Ru2O6}, $C_{m}$ does not approach zero at $T$ $\rightarrow$ 0 like in \ch{H3LiIr2O6}. That indicates system is not gapped. In zero magnetic field, $C_{m}/T$ diverges with $T^{-0.9}$  below about 4 K could be due to vacancy (missing magnetic atom) and quasi-vacancy (spinless impurity on a magnetic site) induced low-energy density of states in the Kitaev Spin Liquid\cite{Perkins2021}. Hence, the ground state will have bound flux in zero magnetic field instead of flux-free sector for pure Kitaev ($C_{m}/T \propto T$). With the application of a magnetic field, ground state becomes flux-free for higher fields\cite{Perkins2021}. Figure \ref{Cm_HLRO96h} shows $C_{m}/T$ as a function of temperature. $C_{m}/T$ is field dependent and follows power law ($\sim$ $T^{1.8}$) variation. These are indication of a gapless excitations from a novel ground state. After obtaining $C_{m}$, we proceeded to compute the magnetic entropy change $\Delta$S$_{m}$ (given by $\Delta$S$_{m} = \int C_{m}/T dT$) for the system at zero field. Figure \ref{Sm_fields} shows magnetic entropy, $\Delta$S$_{m}$ as a function of temperature at various fields: 0-6 Tesla for HLRO. At 100 K, magnetic entropy value $Rln2$ is recovered. Since $Rln2$ is corresponds to $J_{eff}= 1/2$ spins entropy, we claim HLRO is a Ru$^{3+}$ with $J_{eff}= 1/2$. There is a two step entropy release in HLRO like in $\alpha$-\ch{RuCl3} except the first step of the release is not at Rln2/2. $\Delta$S$_{m}$ is field dependent and suppress with increasing field similar like $C_{m}$. We conjecture that the two step entropy release corresponds to the freeing up of two types of quasiparticles of the KQSL; $Z_{2}$ flux at low-$T$ feature and itinerant Majorana Fermions at the high-$T$ feature. However, the ratio of $\Delta$S$_{m}$ at the two plateau is about 1/100 against the expected value of $1/2$, but the ratio of the two corresponding temperature, $T_{L}/T_{H} \sim 0.03$ is consistent with the expected value of 0.03 in the isotropic Kitaev model. So far only $\alpha$-\ch{RuCl3}, a field induced KQSL reveals two step entropy release features as an evidence of $Z_{2}$ flux and itinerant Majorana fermions. There are also distinct differences from expected values: $(i) (1/2)Rln2$ recovery at $T_{H}$ instead of onset of $T_{L}$, (ii) ratio of $T_{L}/T_{H} \sim 0.22$ against the expected value of 0.03 in the isotropic Kitaev model. A small amount of vacancy and quasi-vacancy in HLRO might be affecting
the first plateau in the entropy release due to bound-flux sector excitations
as we have seen low temperature variation of $C_{m}/T \propto T^{-0.9}$ at zero field 
deviates from pure Kitaev limit ($C_{m}/T \propto T$)\cite{Perkins2021}. 
Alternately,
additional couplings such as Heisenberg and off-diagonal exchange terms 
has been shown as a possibility for the unequal entropy 
release~\cite{Yamaji_etal_PRB16, Catuneanu_etal_NPJ_2018, Li_etal_PRR20}.
Given the scalings observed in HLRO, the unequal entropy release 
may be ascribed to vacancy effects primarily.

\section{Nuclear Magnetic Resonance}
\label{D-NMR-HLRO96h}
NMR is a useful local probe of low energy excitations in magnetic insulators. For a recent example of a NMR study on a Kitaev honeycomb system, see Reference\cite{Kitagawa2018}. In our system, the probe nucleus, $^{7}$Li sits at the center of Ru-honeycomb in octahedral environment which leads to zero electric field gradient.

\subsection{$^{7}$Li NMR}

The nucleus $^{7}$Li is an ideal candidate for NMR investigations due to its $I = 3/2$ nuclear spin and 92.6{\%} natural abundance. The gyromagnetic ratio of $^{7}$Li is 16.54 MHz/Tesla. For our NMR study of the \ch{H_{5.9}Li_{0.1}Ru2O6} sample, we utilized a Cu-coil with a frequency range of 90-180 MHz, and the sample's mass was 376 mg. Two types of solid-state-NMR were used for measurement. One is persistent magnet NMR (Fourier Transform NMR) and another is NMR with fixed frequency and variable field (Field Sweep NMR). We exposed the sample to a homogeneous magnetic field of approximately 93.95 kOe (persistent magnet) and excited the target nucleus with a radio frequency (rf) pulse equal to its Larmour frequency ($\omega_{0}$). The spectra were obtained by Fourier transforming the spin-echo after a $\pi$/2$-$$\pi$
pulse sequence. We determined the spin-lattice relaxation time (T$_{1}$) using a saturation recovery
pulse sequence with a $\pi$/2 pulse of 5 ms.

\subsubsection{$^{7}$Li NMR Spectra}

\begin{figure}[!h]
	\centering
	\includegraphics[width=1.0\linewidth]{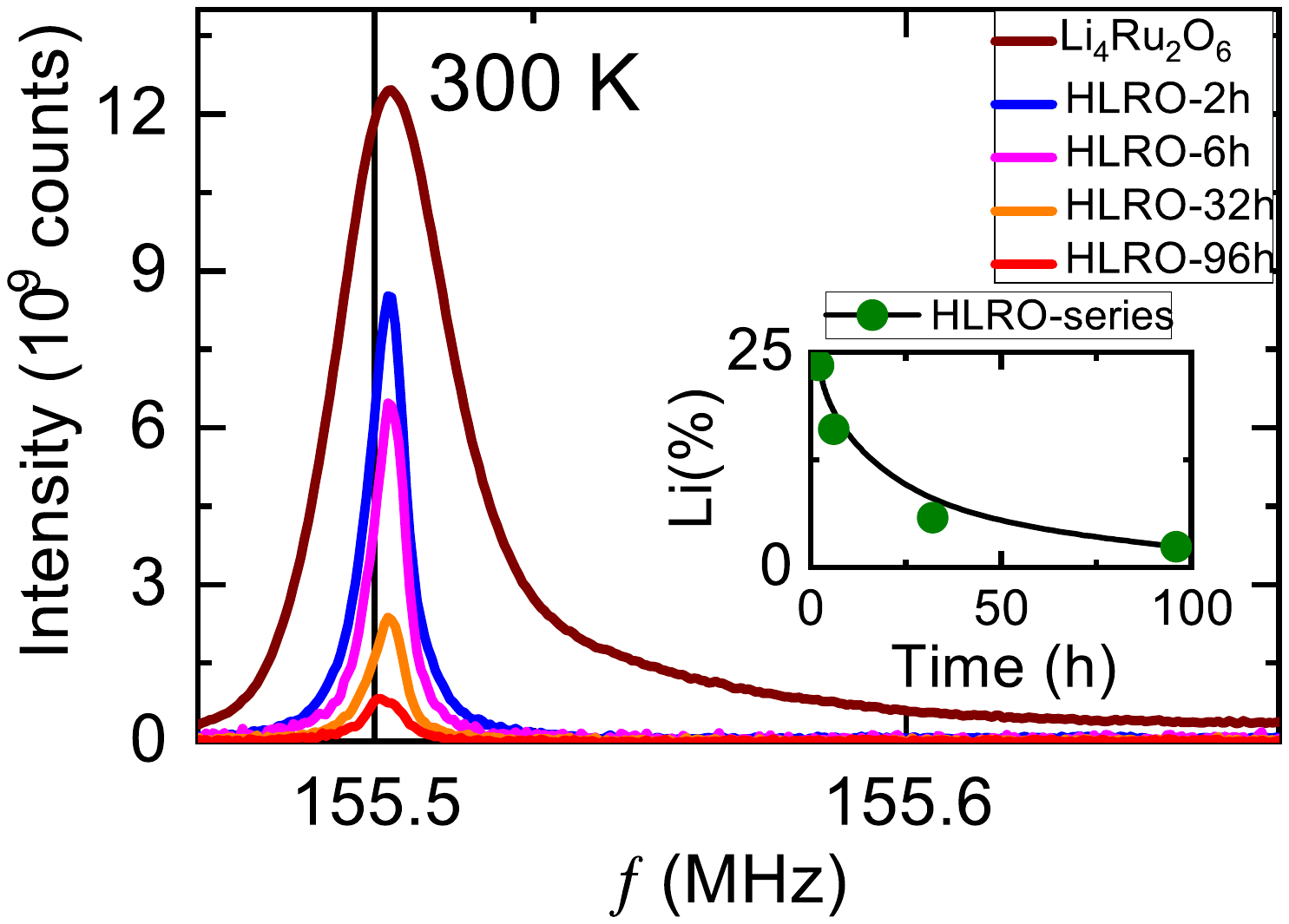}
	\caption{Various $^7$Li NMR Spectra for precursor-\ch{Li4Ru2O6} and sets of HLRO named after intercalation time; HLRO-2h, HLRO-6h, HLRO-32h, HLRO-96h. Inset shows Li content variation as a function of intercalation time.}
	\label{NMRSpectra_setHLRO}
\end{figure}

\paragraph{Li-quantitative analysis} We measured $^{7}$Li NMR spectra at 300 K for various H-intercalated samples (intercalation time: 2, 6, 32, 96 hours) and precursor-\ch{Li2RuO3} (\ch{Li4Ru2O6}) (Figure \ref{NMRSpectra_setHLRO}). 

The solid black line is the reference position at 155.457 MHz. The brown color broad line corresponds to $^{7}$Li NMR spectra of the precursor-\ch{Li4Ru2O6}. There is no shift in spectral line of \ch{Li4Ru2O6}. The possible reason is weak hyperfine coupling although system is magnetic. The y-axis is the absolute intensity scaled by molar mass. In order to compare different spectra, we have kept all technical NMR parameters same for all samples. The blue, pink, yellow, red lines are $^{7}$Li NMR spcetra for HLRO-2h, HLRO-6h, HLRO-32h, HLRO-96h respectively. There are also no shift in NMR spectra, same as precursor-\ch{Li4Ru2O6}. But a narrowing of the NMR line compared to \ch{Li4Ru2O6} is observed. That indicates amount of Li present in HLRO-series is less than that in \ch{Li4Ru2O6}. The absolute intensity gradually decreases with increasing intercalation time of sample. We calculated the Li per formula unit of HLRO-series by the equation;

\begin{equation}
	\frac{I_{abs}(H_{x}Li_{y}Ru_{2}O_{6})}{I_{abs}(Li_{4}Ru_{2}O_{6})} = \frac{y}{4}	
	\label{fcal_HLRO}
\end{equation}              
where ${I_{abs}(H_{x}Li_{y}Ru_{2}O_{6})}$ is the absolute $^{7}$Li NMR spectral intensity of \ch{H_{x}Li_{y}Ru2O6} (HLRO-series) and     ${I_{abs}(Li_{4}Ru_{2}O_{6})}$ is the absolute $^{7}$Li NMR spectral intensity of \ch{Li4Ru2O6}. We calculated absolute $^{7}$Li NMR spectral intensity for all samples: HLRO-2h, HLRO-6h, HLRO-32h, HLRO-96h. The ratio of intensity in percentage are shown in inset of Figure \ref{NMRSpectra_setHLRO}. In HLRO, the percentage of Li is decreasing with increasing intercalation time. Finally, Li per formula unit is calculated by equation \ref{fcal_HLRO} and it is 1, 0.6, 0.2, 0.1 for HLRO-2h, HLRO-6h, HLRO-32h, HLRO-96h respectively. We have XPS evidence of Ru valency of HLRO-96h which is +3. HLRO-96h with Li $= 0.1$ (per formula unit) and Ru$^{3+}$, leads to H$=5.9$ (per formula unit) under charge balance. That leads to the formula \ch{H_{5.9}Li_{0.1}Ru2O6} for HLRO-96h. The compound HLRO-96h or \ch{H_{5.9}Li_{0.1}Ru2O6} is called as HLRO throughout main paper as well as supplementary materials.   

\begin{figure}[!h]
	\centering
	\includegraphics[width=1.0\linewidth]{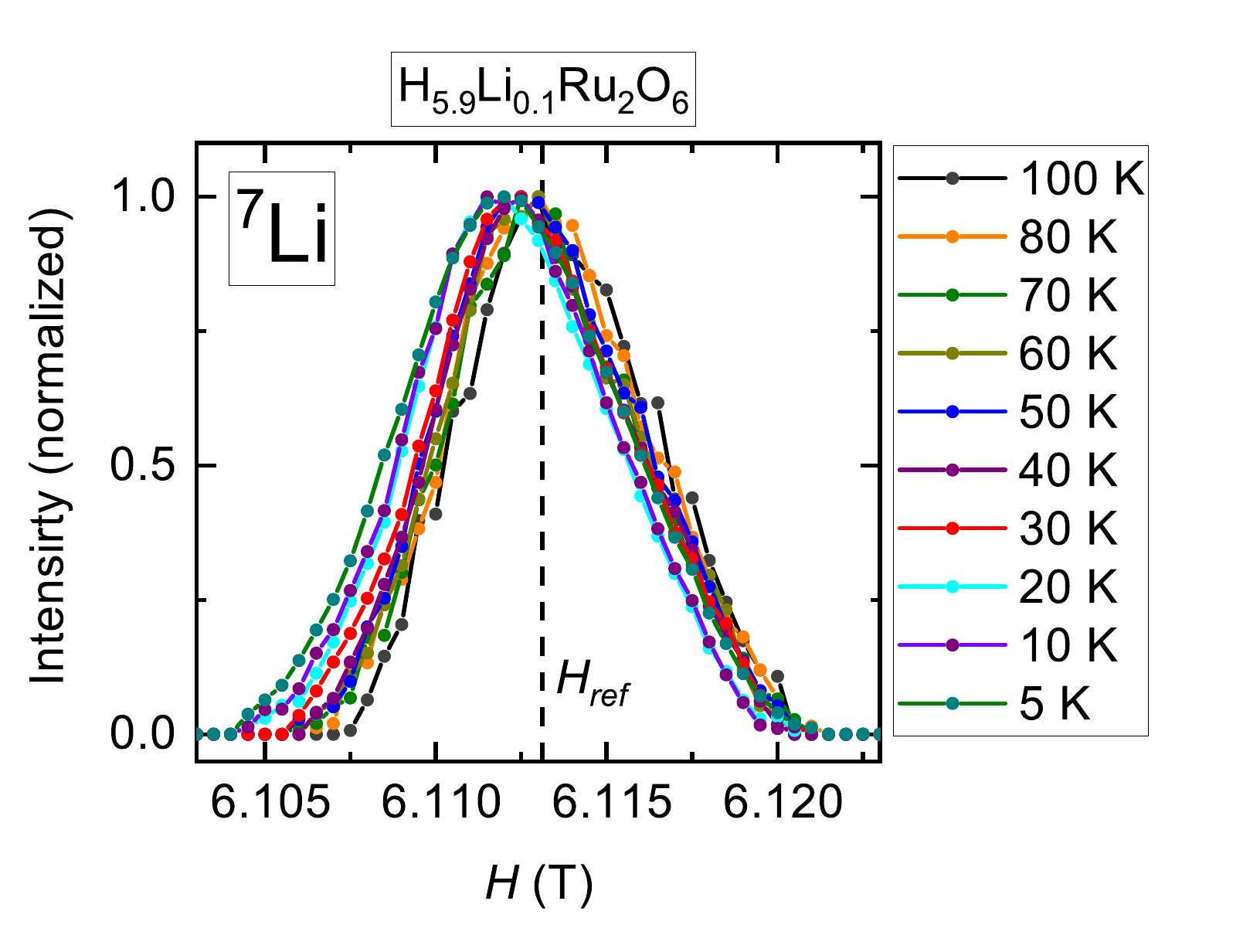}
	\caption{$^7$Li NMR field sweep spectra at 100.29 MHz in the temperature range 5-100 K. the black dotted line is the reference position, H$_{ref}$.}
	\label{HLRO96h_FSSpectra}
\end{figure}

Figure \ref{HLRO96h_FSSpectra} shows $^7$Li NMR field sweep spectra of HLRO at 100.29 MHz in the temperature range 5-100 K. The black dotted line is the reference position, H$_{ref}$. We observe there is almost no shift down to 50 K, then change to non-zero-shift followed by a leveling-off below 20 K. We observe there is almost no change in line broadening down to 10 K and slight broadening at 5 K spectra.

\begin{figure}[!h]
	\centering
	\includegraphics[width=1.0\linewidth]{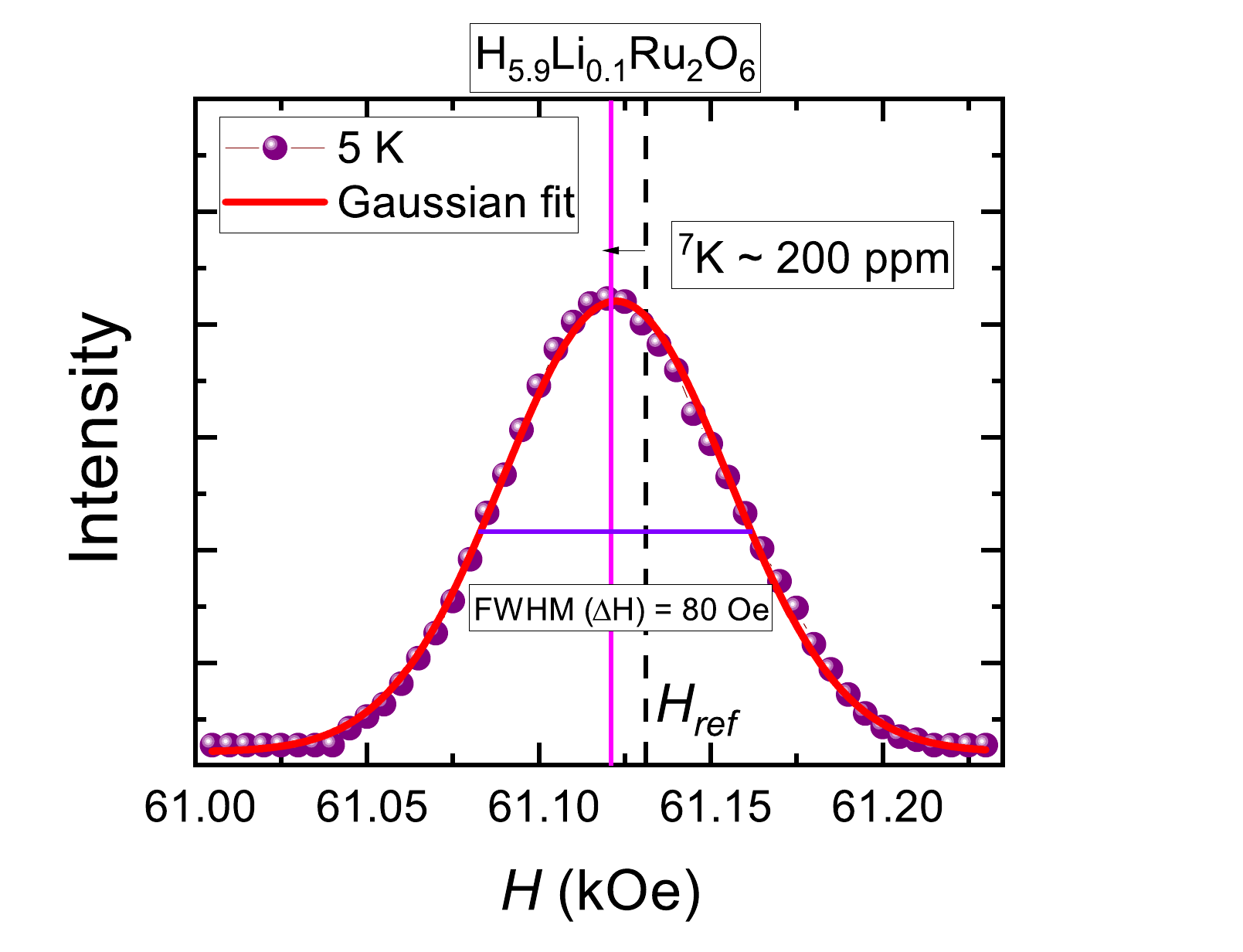}
	\caption{$^7$Li NMR field sweep spectra (frequency = 100.29 MHz) at 5 K. The black dotted line is the reference position, H$_{ref}$.}
	\label{5K_FS_HLRO96h}
\end{figure}

\begin{figure}[!h]
	\centering
	\includegraphics[width=1.0\linewidth]{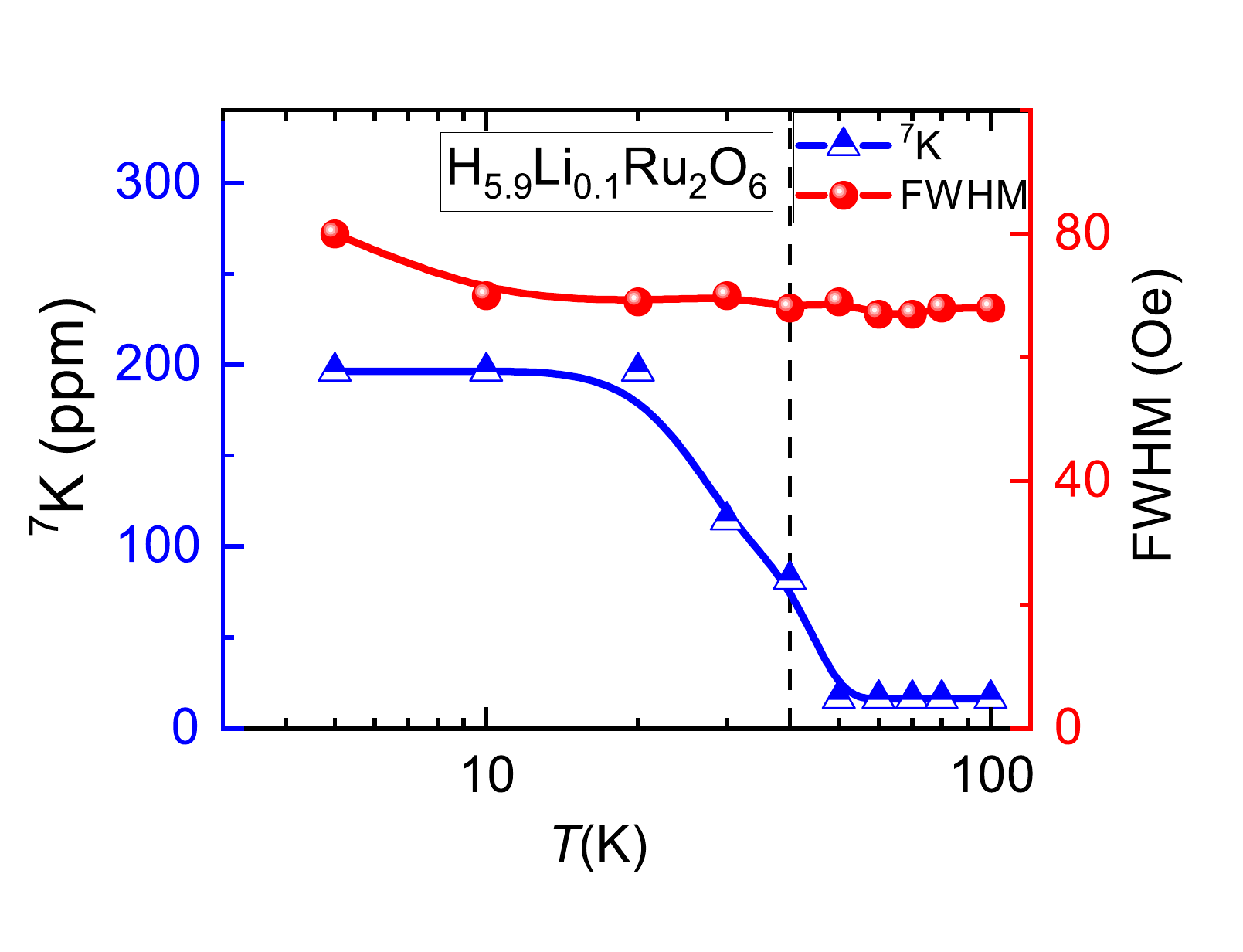}
	\caption{The left y-axis shows variation of NMR Shift, $^{7}$K with temperature,$T$. The right y-axis shows FWHM (full width half maxima) with T. }
	\label{K_HLRO96h}
\end{figure}

\paragraph{NMR Shift, $^{7}$K and FWHM}
We have fitted the spectra to Gaussian function to obtain peak position and FWHM value. We have shown one field sweep spectrum (5 K) fitting in Figure \ref{5K_FS_HLRO96h}. We calculated NMR shift, $^{7}$K $\sim$ 200 ppm and FWHM $\sim$ 80 Oe at 5 K. Figure \ref{K_HLRO96h} (left y-axis) shows variation of NMR Shift, $^{7}$K with temperature, $T$. There is a clear change in $^{7}$K $\sim$ 40 K and then levels off below 20 K. This is indeed consistent with bulk magnetic susceptibility as there was also broad anomaly $\sim$ 40 K. The temperature independent non-zero ($\sim$ 200 ppm) $^{7}$K below 20 K might be a signature of QSL. Figure \ref{K_HLRO96h} (right y-axis) shows FWHM (full width half maxima) variation with temperature. We observe a constant FWHM down to 10 K and then slight increase at 5 K. That indicates internal field distribution is independent of temperature like KQSL, \ch{H3LiIr2O6} \cite{Kitagawa2018}. 
     
\subsubsection{$^{7}$Li Spin-lattice relaxation rate, 1/T$_{1}$}

\begin{figure}[!h]
	\centering
	\includegraphics[width=1.0\linewidth]{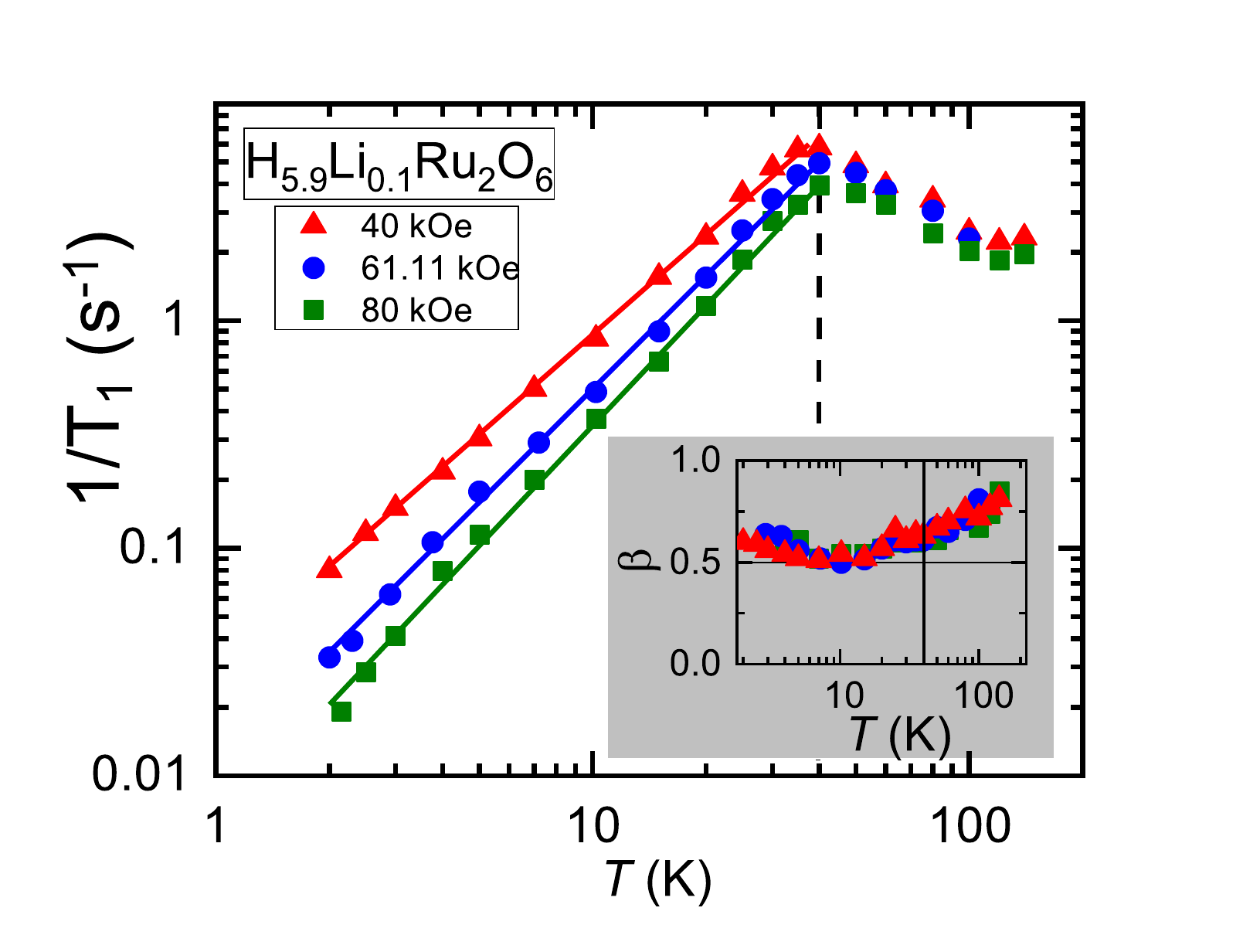}
	\caption{The variation of $^7$Li 1/T$_{1}$ as a function of temperature for \ch{H_{5.9}Li_{0.1}Ru2O6}
		at different applied magnetic fields: 40 kOe (red triangle), 61.11 kOe (blue circle),
		80 kOe (green square). (Inset) Stretch exponent, $\beta$ vs Temperature, $T$.}
	\label{T1_Li_HLRO96h}
\end{figure}

The Figure \ref{T1_Li_HLRO96h} shows the variation of 1/T$_{1}$ as a function of temperature for \ch{H_{5.9}Li_{0.1}Ru2O6} at different applied magnetic fields (40 kOe, 61.11 kOe and 80 kOe) in the temperature range 2-150 K. The anomaly $\sim$ 40 K has been also observed in other measurements like magnetization and heat capacity. Below 40 K, 1/T$_{1}$ exhibits a power law variation with temperature. While the anomaly at $\sim$ 40 K is not likely due to long range ordering (LRO) or glassy magnetism (no evidence in $\chi(T)$ or $C_{m}(T)$), the power law suggests gapless excitations. The spin lattice relaxation rate, 1/T$_{1}$ suppressed below 40 K with increasing applied field. A similar effect was seen in \ch{H3LiIr2O6} and was attributed to field dependent density of states (DOS).

\subsection{$^{1}$H NMR}

The nucleus $^{1}$H is an ideal candidate for NMR investigations due to its $I = 1/2$ nuclear spin and
99.98{\%} natural abundance. The gyromagnetic ratio of $^{1}$H is 42.575 MHz/Tesla. In \ch{H_{5.9}Li_{0.1}Ru2O6}, $^{1}$H occupied inter layer as well in-plane position. The spectra were obtained by field sweep method at MPI-CPFS, dresden. We determined the spin-lattice relaxation time (T$_{1}$) using a saturation recovery pulse sequence with a $\pi$/2 pulse of 3 ms.

\subsubsection{$^{1}$H NMR Spectra}

\begin{figure}[!h]
	\centering
	\includegraphics[width=1.0\linewidth]{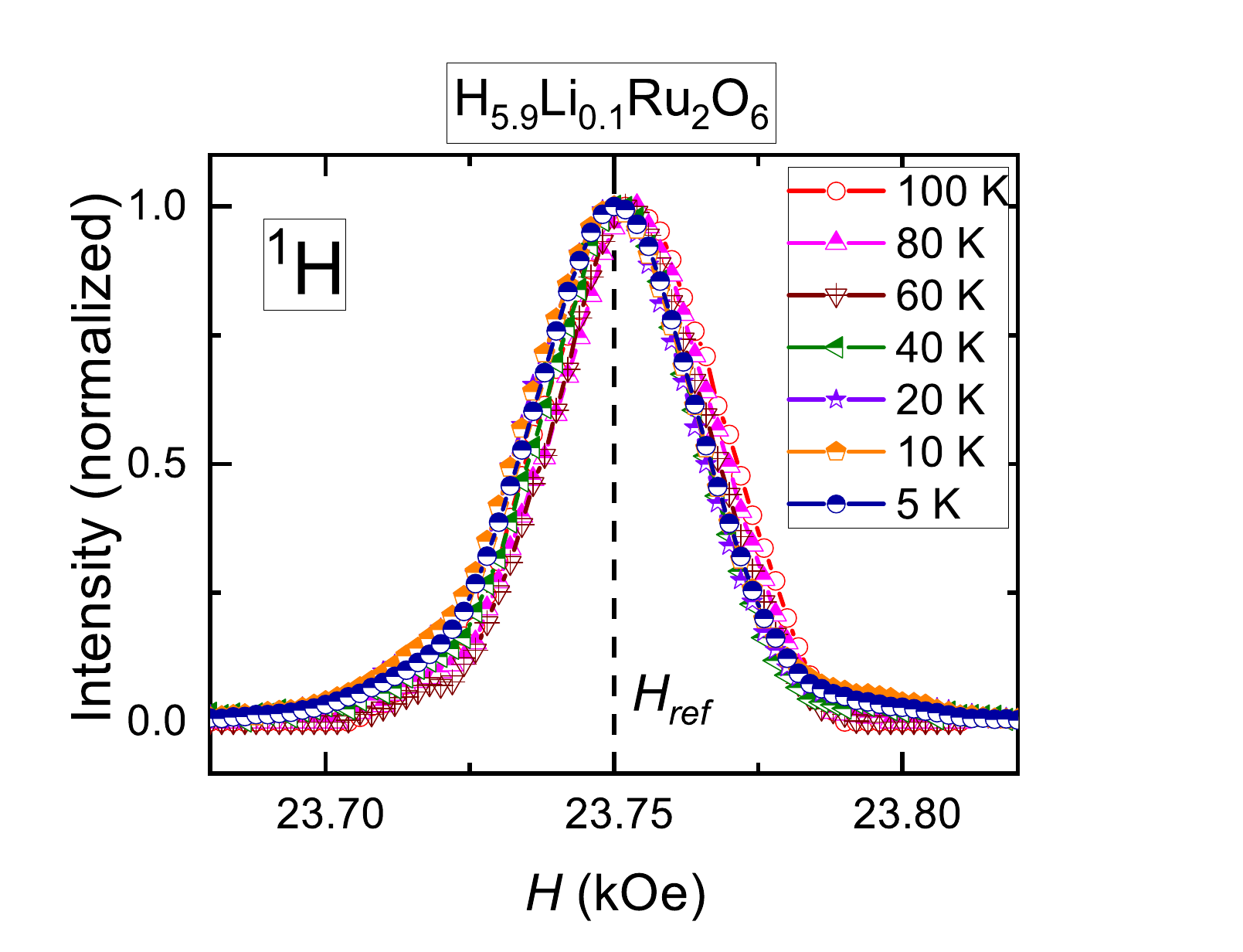}
	\caption{$^{1}$H NMR field sweep spectra in the temperature range 5-100 K for \ch{H_{5.9}Li_{0.1}Ru2O6}. The black dotted line is the reference position, H$_{ref}$.}
	\label{Spectra_H_HLRO96h}
\end{figure}

Figure \ref{Spectra_H_HLRO96h} shows $^{1}$H NMR field sweep spectra in the temperature range 5-100 K. We observe that there is no shift and also independent of temperature. The line width, FWHM is also constant. That indicates internal field distribution is same. We have seen such scenario in KQSL, \ch{H3LiIr2O6}.

\subsubsection{$^{1}$H Spin-lattice relaxation rate, 1/T$_{1}$}

\begin{figure}[!h]
	\centering
	\includegraphics[width=1.0\linewidth]{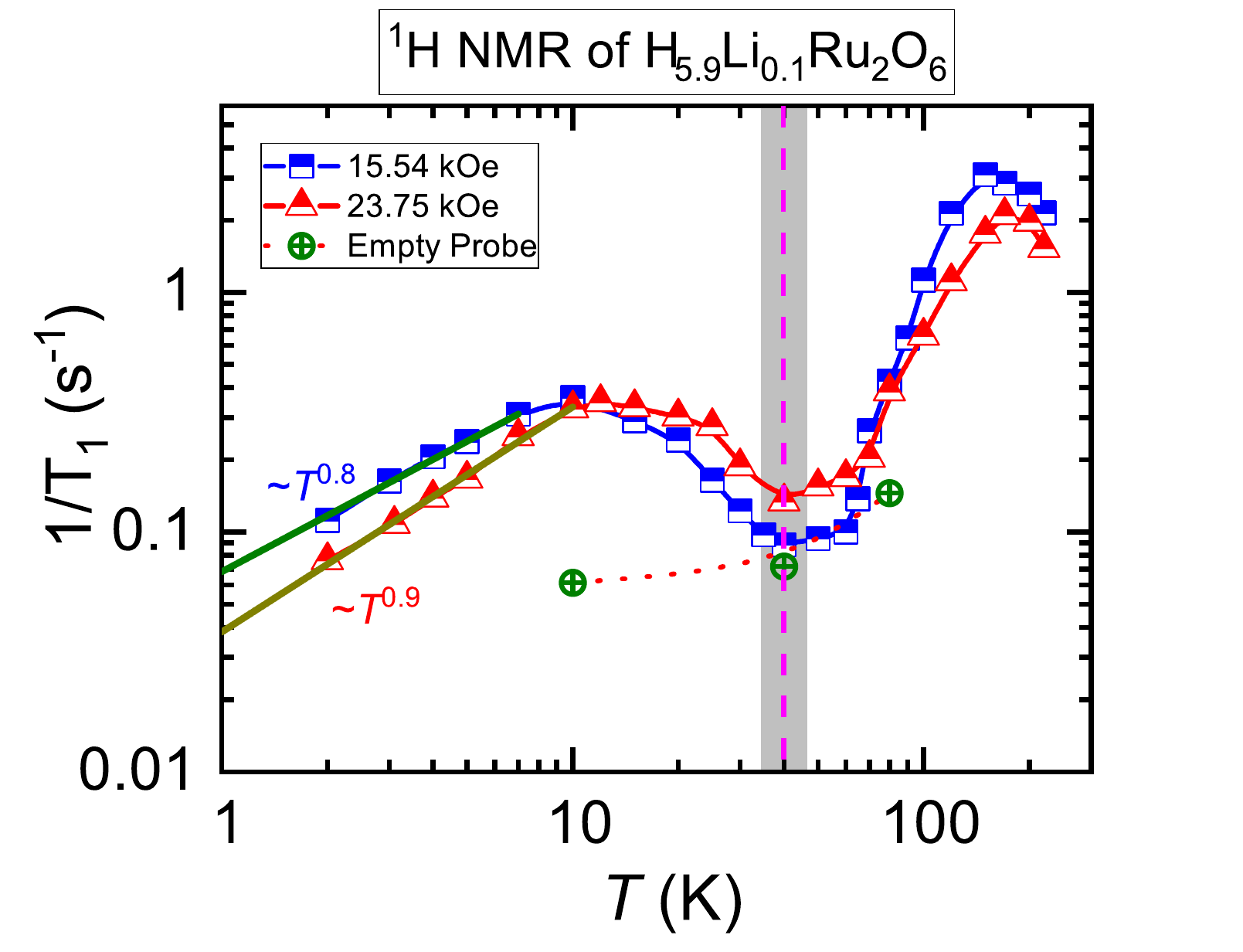}
	\caption{The Variation of $^{1}$H 1/T$_{1}$ as a function of temperature for \ch{H_{5.9}Li_{0.1}Ru2O6} at 15.54 kOe (blue square) and 23.75 kOe (red triangle). The variation
	of 1/T$_{1}$ for probe is shown by green circle.}
	\label{T1_H_HLRO96h}
\end{figure}

Figure \ref{T1_H_HLRO96h} shows variation of spin-lattice relaxation rate, 1/T$_{1}$ as a function of temperature for \ch{H_{5.9}Li_{0.1}Ru2O6} at 15.54 kOe and 23.75 kOe and also for probe. We observe 1/T$_{1}$ has a peak around 150 K and then a minimum $\sim$ 40 K. Below 40 K, it has another broad anomaly around 10 K followed by a power law variation. The variation of 1/T$_{1}$ for 15.54 kOe (blue square) and 23.75 kOe (red triangle) are similar but quantitative values are different. We observe 1/T$_{1}$ value for empty probe is close to
\ch{H_{5.9}Li_{0.1}Ru2O6} down to 40 K. It looks
1/T$_{1}$ for probe also varies with temperature. Commenting on 1/T$_{1}$ data above 40 K is risky
because separation of probe contribution from measured sample data looks tricky. However,
T$_{1}$ at 10 K for probe and HLRO are factor of 5 difference with constant stretch exponent, $\beta \sim 0.75$. Below 10 K, power law
variation of spin lattice relaxation rate ($\sim$ $T^{0.8-0.9}$) indicates gapless excitation.

\section{Muon Spin Relaxation ($\mu$SR)} 

\paragraph{Zero field $\mu$SR at 1.56 K} 
The Oscillations in ZF-$\mu$SR are key signature for magnetically ordered system below transition temperature \cite{AVM2021}. Magnetic spins get order and muon spins precess
about static internal field. In presence of static spins, $1/3$ of the local magnetic field components
pointing parallel to the initial muon spin (and are therefore unchanged by the magnetic
field) which gives rise to $1/3$-tail in muon asymmetry and $2/3$ of the magnetic field components
in a perpendicular direction which gives rise to oscillations in asymmetry. If spins are
dynamic and internal local field fluctuate continuously, exponential decay in asymmetry is expected
without any $1/3$-tail. We performed zero field $\mu$SR at $1.56$ K. We do not observe any
oscillation. That indicates there is no long range ordering. We also do not observe any $1/3$-tails.
That indicates absence of static moments. Fitting the data to single exponential ($A(t)= A(0)exp(- \lambda t)$) yields $\lambda = 0.7$ MHz, does not fit well. Then, we tried with stretch exponential ($A(t) = A(0)exp(- (\lambda t)^{\beta})$) (fitting parameter $\lambda = 0.68$ MHz and $\beta = 0.6$) which is also does not fit either, not fitting short time asymmetry decay properly. Finally, fitting to the data with double exponential function ($A(t) = A(0)[A_{1}exp(- \lambda_{1} t) + A_{2}exp(- \lambda_{2}t) + (1 - (A_{1} + A_{2}))]$) looks the best fit which yields $A_{1} = 0.256$ with $\lambda_{1} = 19.8$ MHz and $A_{2} = 0.70$
with $\lambda_{2} = 0.48$ MHz.

  \begin{figure}[!h]
  	\centering
  	\includegraphics[width=1.0\linewidth]{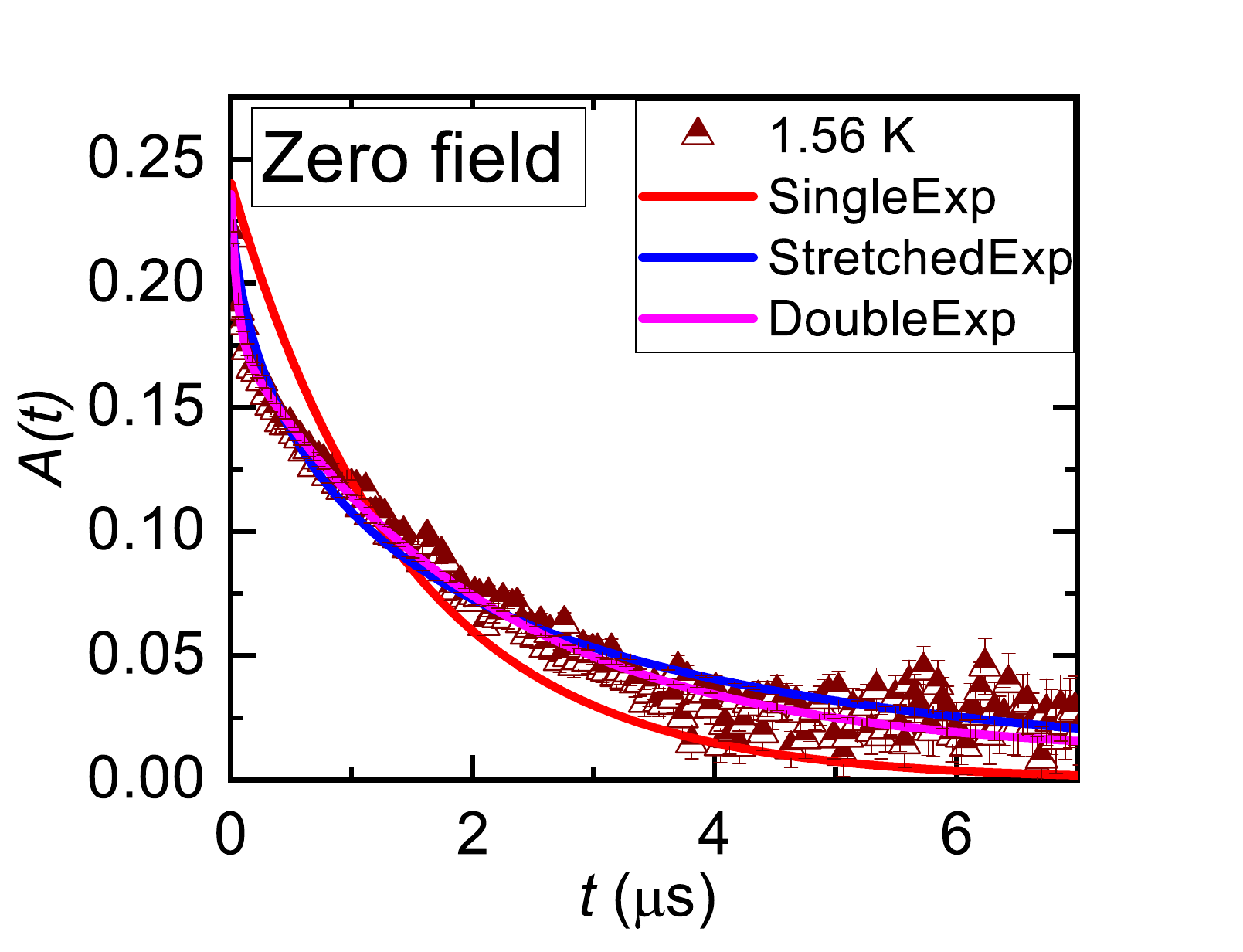}
  	\caption{Muon asymmetry as a function of decay time for zero field (ZF) at
  		1.56 K.}
  	\label{1.56K_ZF}
  \end{figure}    
\paragraph{100 Oe-LF $\mu$SR (1.56-300 K)}

\begin{figure}[!h]
	\centering
	\includegraphics[width=1.0\linewidth]{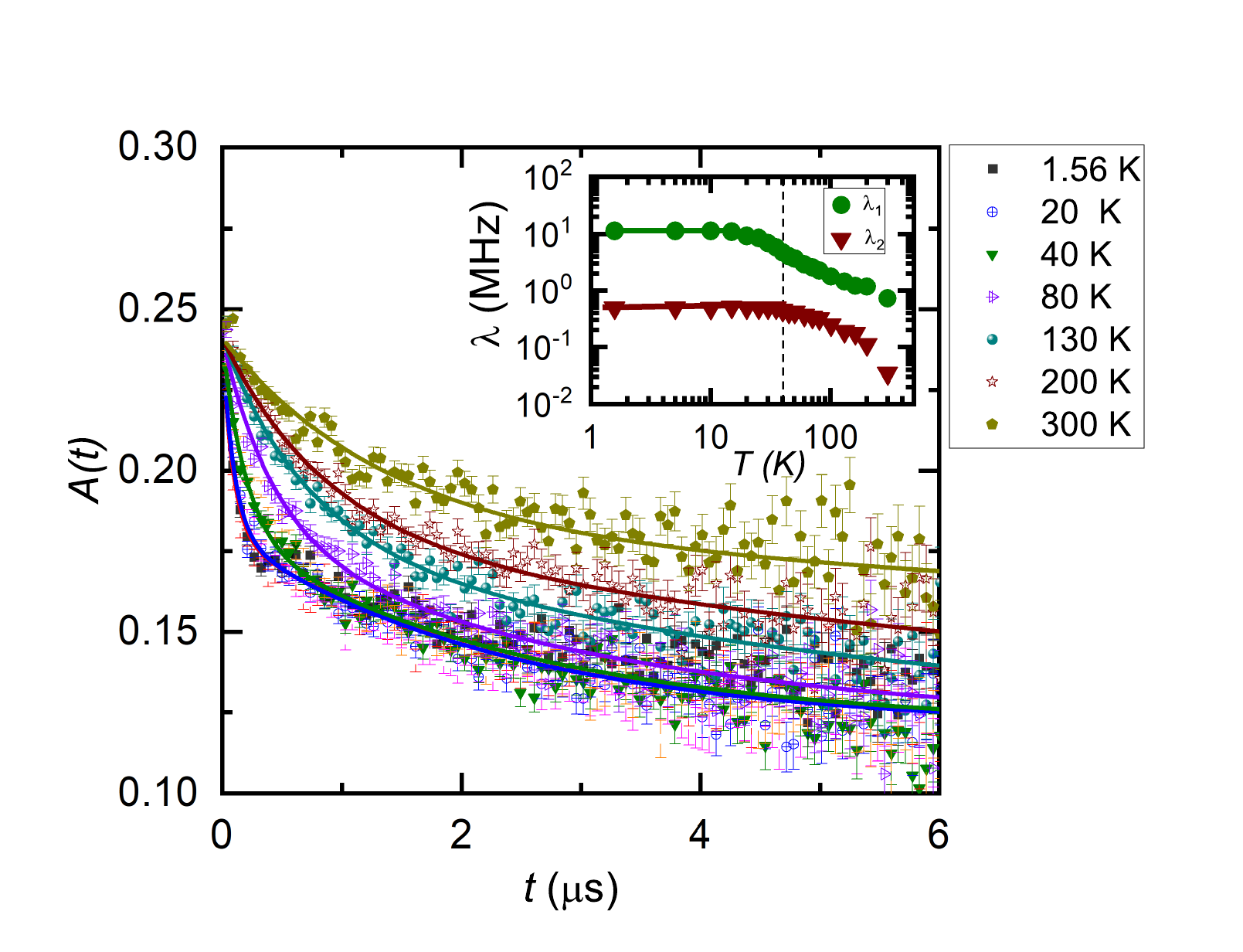}
	\caption{Muon asymmetry as a function of decay for 100 Oe longitudinal field in the temperature range: 1.56 K-300 K. (Inset)  Muon relaxation rate $\lambda$ as a function of temperature. Two components of relaxation, $\lambda_{1}$ (olive circle) and $\lambda_{2}$ (wine circle) with $A_{1}$ : $A_{2}$ = 1:1 are observed.  }
	\label{100G_LF}
\end{figure}

Zero field $\mu$SR measurements include contribution from nuclear and magnetic/electronic part. We applied a longitudinal field (LF) of 100 Oe which should remove the depolarization due to nuclear moments. Data were taken as a function of temperature between 1.56 K and 300 K in the longitudinal field of 100 Oe (Figure \ref{100G_LF}). The relaxation/ depolarization thus observed should be only due to the electronic moments. We observe data between 1.56 K and 20 K overlaps indication of same relaxation rate $\lambda$. Above 20 K, data gradually changes its slope and exponential decay nature noticed even at 300 K.       

\begin{figure}[!h]
	\centering
	\includegraphics[width=0.9\linewidth]{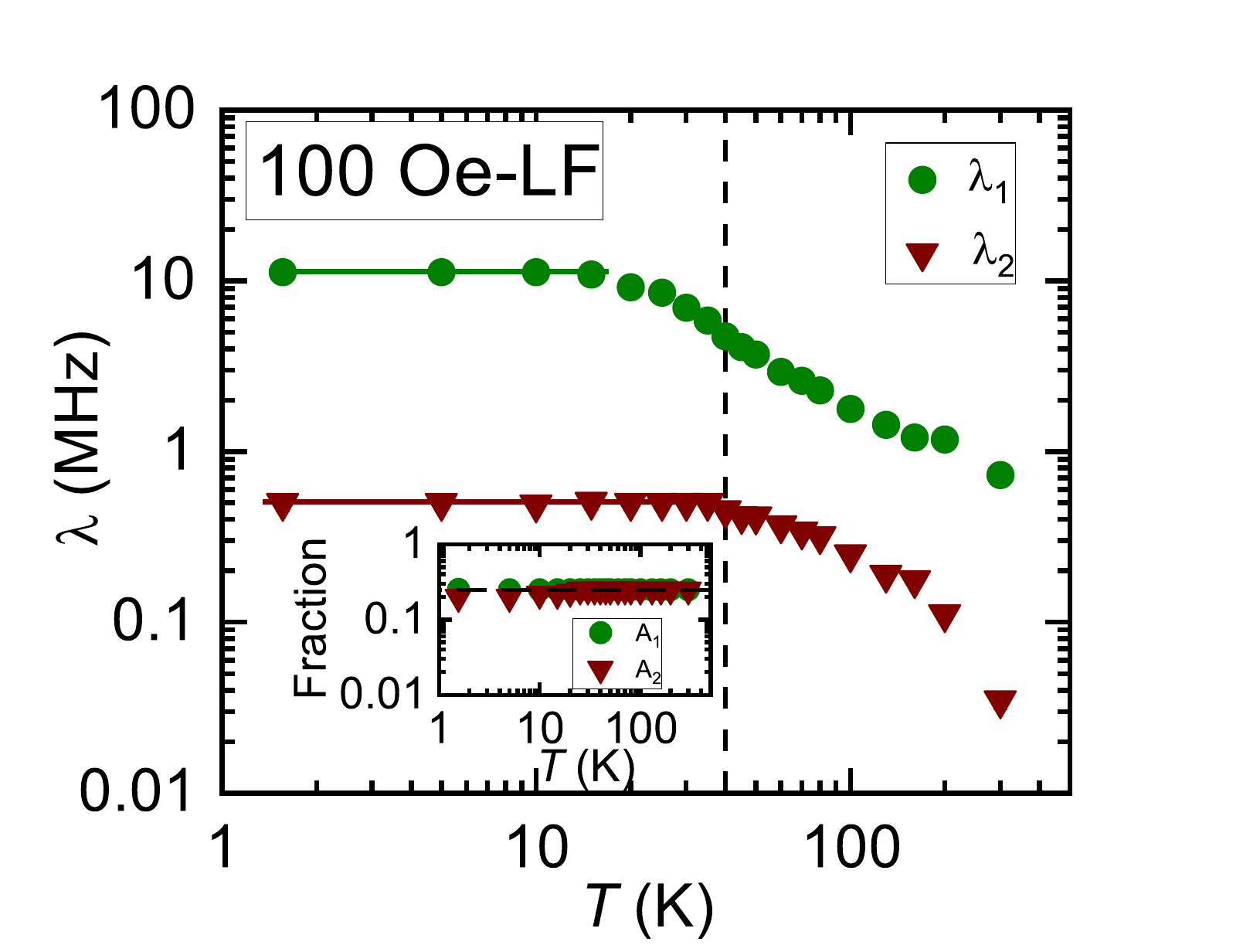}
	\caption{Muon relaxation rate, $\lambda$ as a function of temperature. Two components of relaxation, $\lambda_{1}$ (olive circle) and $\lambda_{2}$ (wine circle) with $A_{1}$ : $A_{2}$ = 1:1 are observed.  }
	\label{Lambda_T}
\end{figure}

We found the muon relaxation could be best fit with a double exponential in addition to a constant as in H$_{3}$LiIr$_{2}$O$_{6}$. Figure \ref{Lambda_T} shows variation of $\lambda_{1}$ and $\lambda_{2}$ as function of temperature. Two components of relaxation, $\lambda_1$ and $\lambda_2$ with equal relative weights ($A_{1}$ : $A_{2} = $ 1:1) are observed. Muon spin relaxation rate, $\lambda$ gradually increasing with lowering the temperature and there is a change in dynamics around $\sim$ 40 K followed by a leveling-off below 20 K. Features at 40 K have been also observed in other bulk and NMR local probes which is a possible indication of a QSL cross over point. The flattening/saturation of $\lambda$ at low-$T$ indicates spin remains dynamic down to low-T. The weights ($A_{1}$ and $A_{2}$) are temperature-independent (see Figure \ref{Lambda_T} (Inset)). That indicates two muon stopping sites are present. Muon ($\mu^{+}$) usually stops $\sim$ 1 {\AA} away from Oxygen ($O$). Possibly there is two $O^{2-}$ sites crystallographically in a given unit cell. These results are similar to those of H$_{3}$LiIr$_{2}$O$_{6}$ except that the flattening/saturation of $\lambda$ sets in at 10 times the temperature in H$_{3}$LiIr$_{2}$O$_{6}$ \cite{Yang2024}.

\paragraph{LF measurements}
\begin{figure}[!h]
	\centering
	\includegraphics[width=1.0\linewidth]{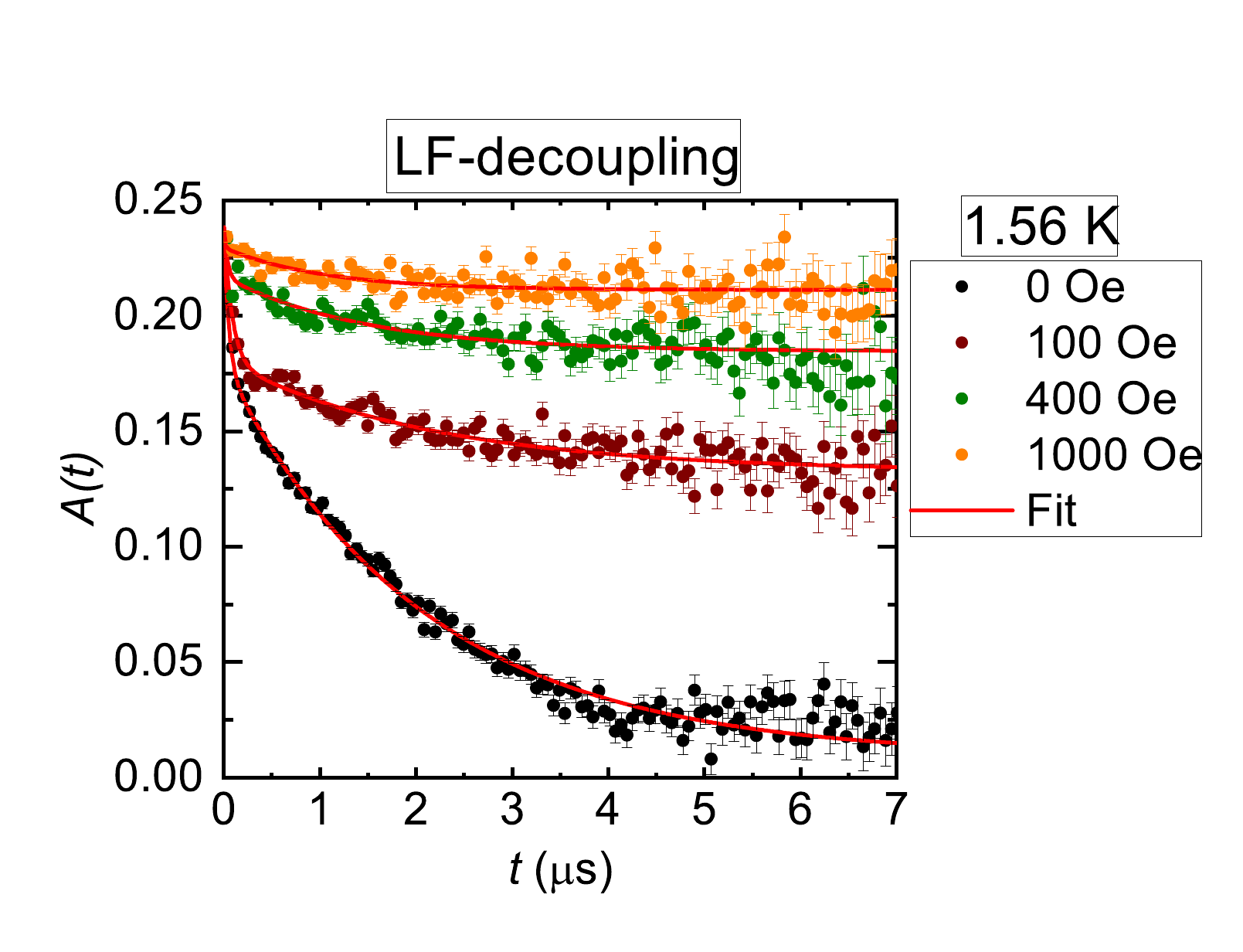}
	\caption{Muon asymmetry as a function of decay time at 1.56 K for various longitudinal fields (GPS data): 0 Oe, 100 Oe, 400 Oe, 1000 Oe.}
	\label{1.56K_LF}
\end{figure}

\begin{figure}[!h]
	\centering
	\includegraphics[width=1.0\linewidth]{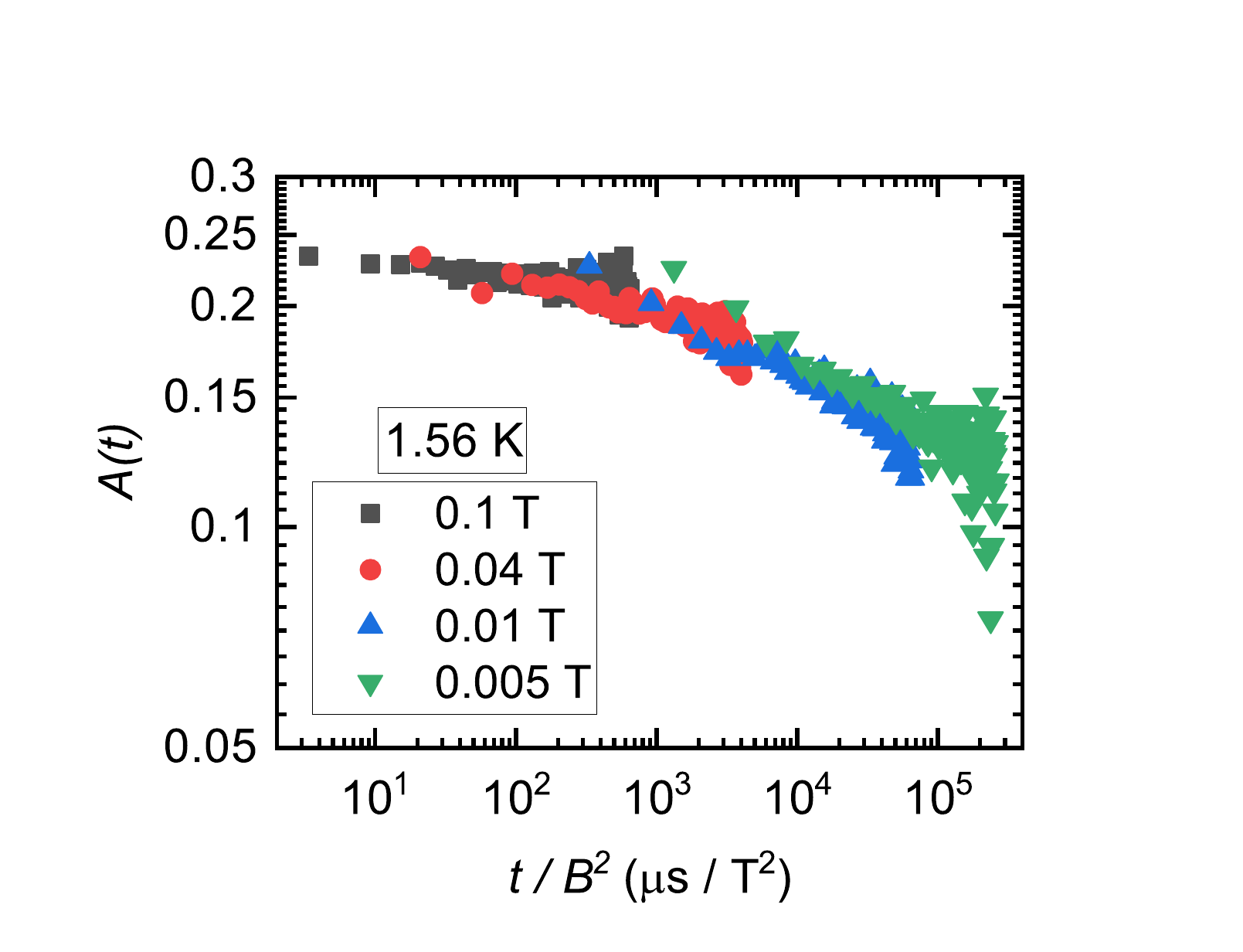}
	\caption{Time-field scaling of LF-$\mu$SR spectra at 1.56 K. }
	\label{timefield_HLRO96h}
\end{figure}

\begin{figure}[!h]
	\centering
	\includegraphics[width=0.9\linewidth]{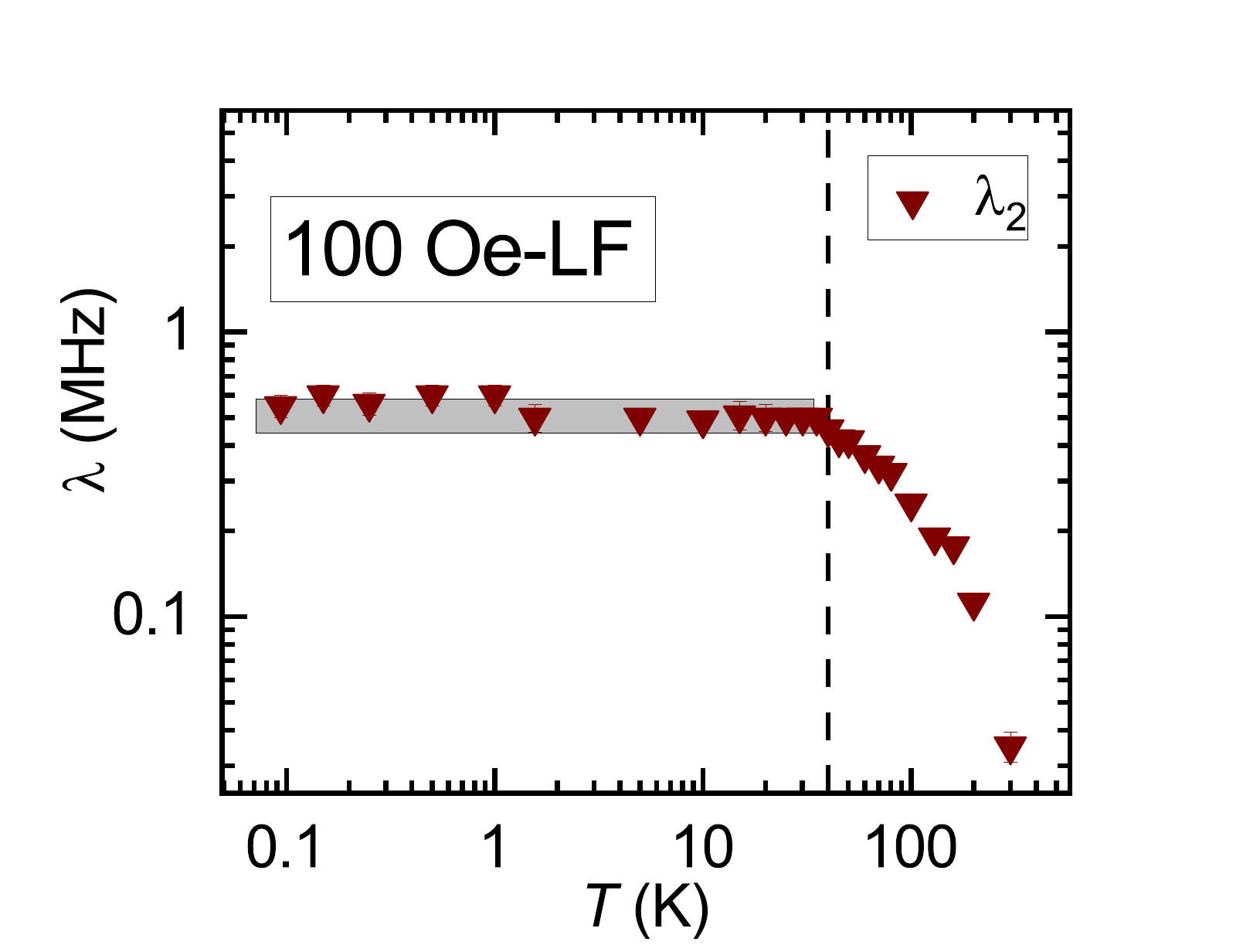}
	\caption{Muon relaxation rate, $\lambda$  (slower component) as a function of temperature.  }
	\label{Lambda2_T_84mK}
\end{figure}

\begin{figure}[!h]
	\centering
	\includegraphics[width=1.0\linewidth]{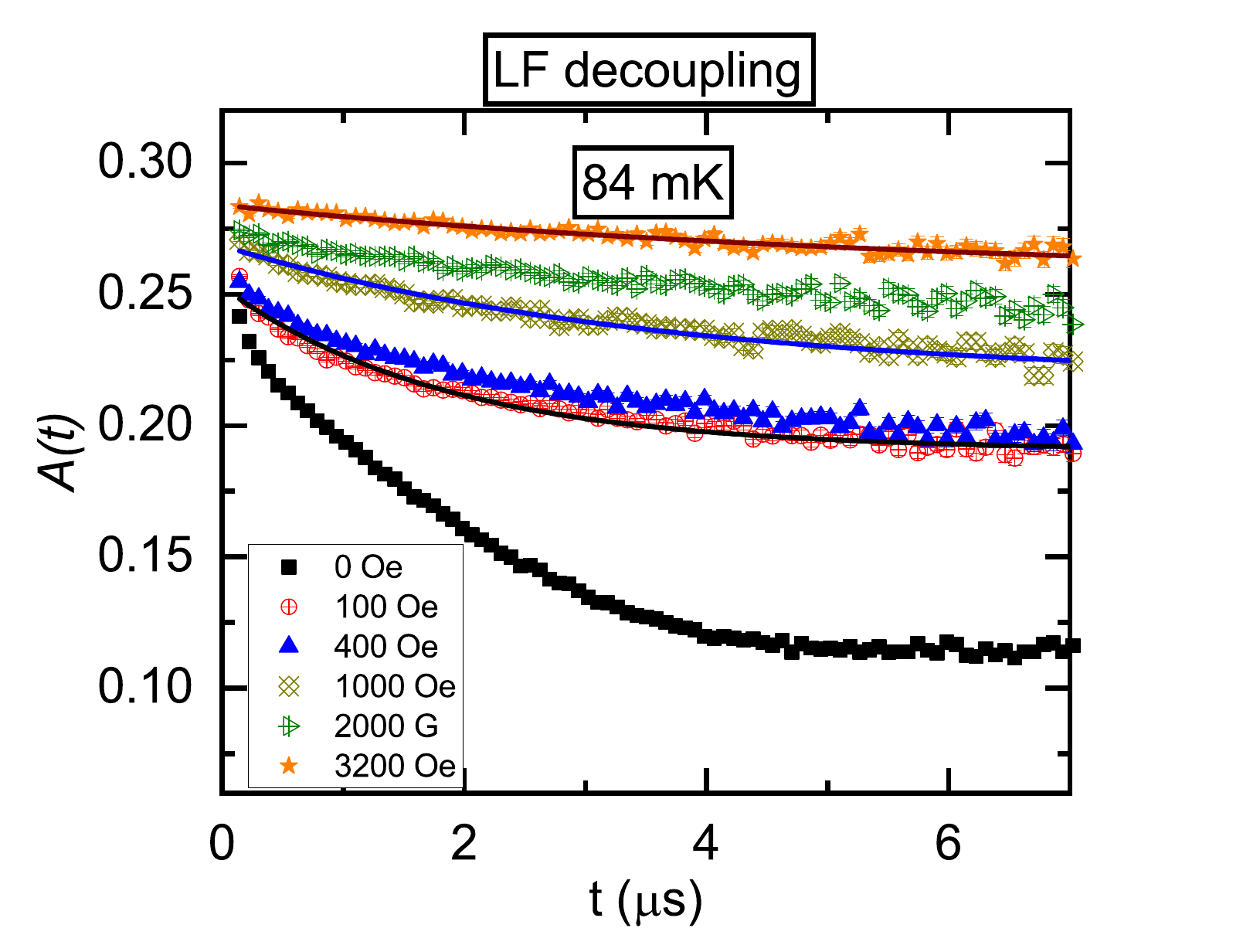}
	\caption{Muon asymmetry as a function of decay time at 84 mK for various longitudinal fields (ISIS data): 0-3200 Oe.}
	\label{LF_asym_HLRO}
\end{figure}

\begin{figure}[!h]
	\centering
	\includegraphics[width=1.0\linewidth]{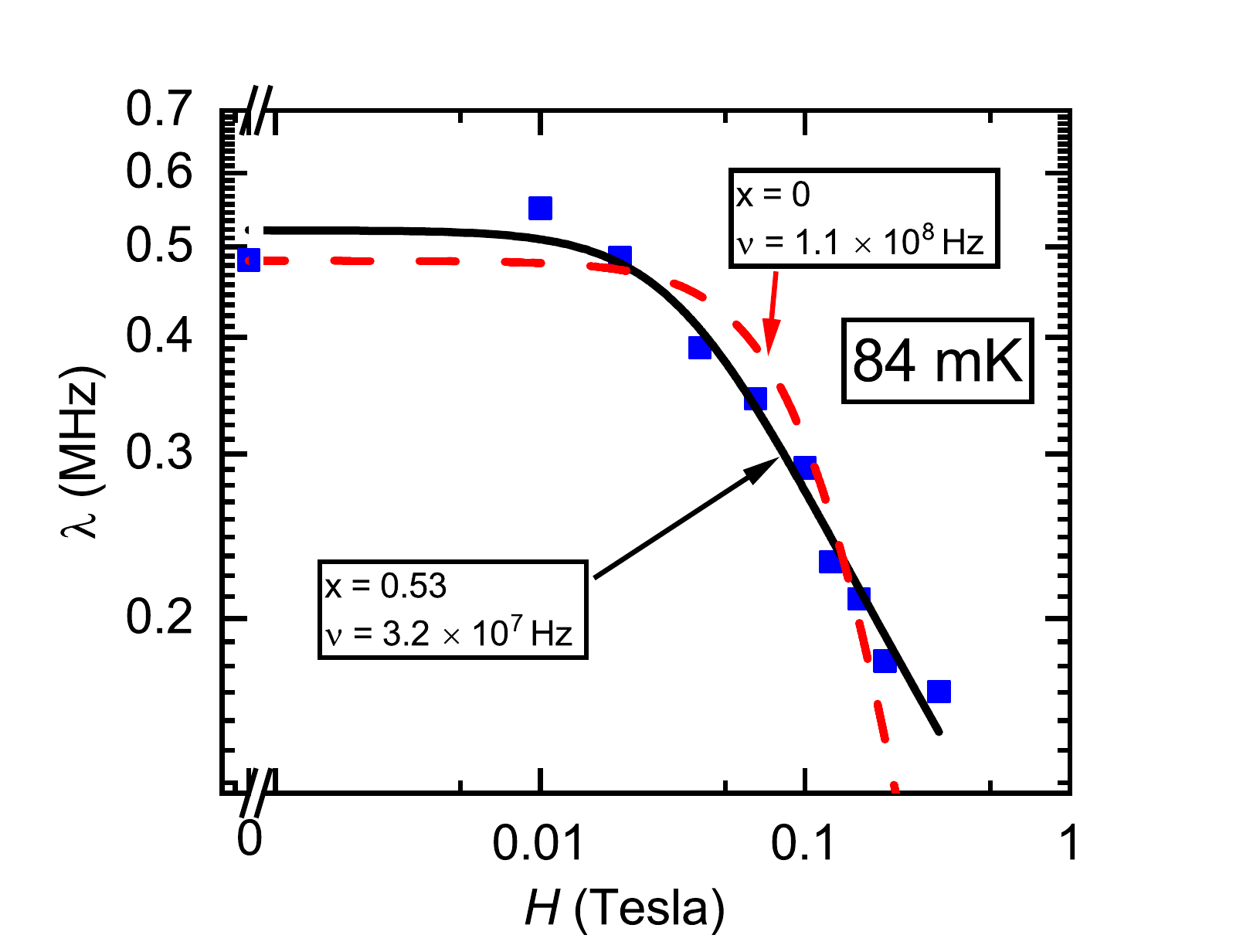}
	\caption{Muon spin relaxation rate, $\lambda$ with longitudinal field, H at 84 mK.}
	\label{lamda_H_HLRO}
\end{figure}

Longitudinal field (LF) measurements are useful to differentiate between static and dynamic magnetism. If the magnetism is static, a longitudinal field with a strength 10 times greater than the field distribution would effectively decouple the muons. On the other hand, if the magnetism is dynamic, a longitudinal field 50 times stronger than the field distribution would be necessary. The nuclear field is $\sim$ 2-3 Oe and hence 50 Oe longitudinal field is enough to remove nuclear contribution. We applied 100-1000 Oe longitudinal field at 1.56 K to differentiate between static and dynamic spins (Figure \ref{1.56K_LF}). In HLRO, ZF/LF-muon asymmetry was well fitted with two exponential plus a constant rather than the usual single exponential \cite{SKundu2020}.  In our case, it probably arises from two muon stopping sites. There is very little decoupling for the first component of the exponential decay with application of 1000 Oe field. This indicates that the spins are dynamic. The second component corresponds to a much slower decay and we have tracked this variation with much higher statistics at the ISIS MUSR beamline down to 84 mK. In the ISIS measurements, the faster component ($\lambda \sim$ 10 MHz) could not be tracked down to 84 mK due to limitations associated with data collection for the pulsed muon source there. No ordering is observed and $\lambda$ (slower component) continues to remain constant with temperature below 1.56 K and down to 84 mK (Figure \ref{Lambda2_T_84mK}) which is consistent with other experiments. From the data in figure \ref{LF_asym_HLRO}, it can be seen that even in a large field of 3200 G, the muons are still not decoupled from the internal fields. This indicates that the moments remain dynamic down to the lowest temperatures studied (84 mK). These are typical signatures seen in quantum spin liquid materials. The LF data were fit to an exponential in addition to a constant background. The muon depolarization rate thus obtained is plotted as a function of the magnetic field in Figure \ref{lamda_H_HLRO}. 

At very high fields one expects muons to totally decouple
from the internal fields resulting in no relaxation.
But as seen in Figure \ref{LF_asym_HLRO}, a field of 3200 G is not at all sufficient to make $\lambda$ near negligible. Following the analysis of the field dependence of $\lambda$ as in Ref. \cite{YMGO}, we fit the
data to the following equation:

\begin{equation}
	\lambda (H) = 2 \delta ^2 \tau ^x \int_{0}^{\infty} t^{-x} exp(- \nu t) Cos (\gamma_{\mu} H t) dt
	\label{lamda_H}
\end{equation}

  where $\nu$ is the fluctuation frequency of local moments and
$\delta$ is the distribution width of the local magnetic fields.
The muon gyromagnetic ratio is 
$\gamma_{\mu}$ = 2$\pi$ x 135.5342
MHz/T. A fit with $x = 0$ (red  dashed line in Figure \ref{lamda_H_HLRO})
which implies an exponential auto-correlation function
$S(t) \sim exp(-\nu t)$ does not fit the data well and rather
$S(t) \sim (\tau/t)^x exp(- \nu t)$ is needed to fit the data. The black solid curve is a fit to Equation \ref{lamda_H} and gives $x = 0.53$ and $\nu = 3.2 \times 10^7$ Hz. $\tau$ is the early time cut-off and is fixed
to 1 ps. This result implies the presence of long-time spin correlations but without any static order; a hallmark of spin liquids. The LF dependence shows a striking resemblance to that observed in \ch{YbMgGaO4} \cite{YMGO} and \ch{Sr3CuSb2O9} \cite{SKundu2020}, characterized by a local moment fluctuation frequency of approximately 32 MHz and the existence of spin correlations over long-time. The qualitative and quantitative results obtained through $\mu$SR analysis of HLRO align with those observed in other quantum spin liquid (QSL) systems\cite{SKundu2020}. Long time correlation in KQSL through LF-$\mu$SR studies is first observed in HLRO.
 Further, Figure \ref{timefield_HLRO96h} shows time-field scaling of LF-$\mu$SR spectra at 1.56 K. We feel that there is some connection between field dependent DOS (Density of States) and this scaling. We observed field dependent NMR-1/T$_{1}$ and $C_{m}$ as an indication of field dependent DOS. We also found time-field scaling in the form $A(t) \sim t/H^{\alpha}$ with $\alpha=2$. We suggest that this time-field scaling is connected to field dependent DOS as seen in \ch{H3LiIr2O6} \cite{Yang2024}.

\section{Model of Density of States (DOE)}
Heat capacity can be written as
\begin{equation}
	\frac{C_m}{T} = \frac{1}{T} \int E D(E) \frac{df(T,E)}{dT}dE 
	\label{C_theory}
\end{equation}

where $f(T,E)$ is Fermi distribution function and can be expressed as 

\begin{equation}
	f(T,E) = \frac{1}{exp(E/k_{B}T) +1}
\end{equation}
 
For scaling, lets assume $E^{\prime}=E/k_{B}T$.

Hence, Equation \ref{C_theory} would be
 
\begin{equation}
	\frac{C_m}{T} = k_{B}^{2} \int  D(E^{\prime}k_{B}T) \frac{{E^{\prime}}^{2}  exp(E^{\prime})} {(exp(E^{\prime}) +1)^{2}} dE^{\prime} 
	\label{C_theory_1}
\end{equation}

For magnetic field, $B=0$ if we model 
$D(E,0) = \Gamma |E|^{-\eta}$ (where $\Gamma$ is a 
constant with appropriate dimensions for $\eta= 0,1,2,\ldots$)

then, 
\begin{equation}
	\frac{C_m}{T} \propto |T|^{-\eta}
\end{equation}

Following along the lines of
Ref.~\cite{Kitagawa2018}, for $B\ne0$, we model 
$D(E,B)$ as:

\begin{equation}
 D(E,B)= \Gamma \frac{|E|^{\gamma}}{(\alpha \mu_B B)^{\gamma + \eta}}
 \end{equation}
where $\gamma$ is an scaling exponent.

then, 

\begin{equation}
	B^{\eta}\left(\frac{C_m}{T}\right) \propto \left(\frac{T}{B}\right)^{\gamma}
\end{equation}

For $\eta=0.5$ and $\gamma=1$, we get $\frac{C_m}{T} \propto |T|^{-0.5}$ and 
$D(E,0) \propto |E|^{-0.5}$ for $B=0$. The scaling would be $B^{0.5}\left(\frac{C_m}{T}\right) \propto \left(\frac{T}{B}\right)$ and $D(E,B\ne0)= \Gamma \frac{|E|^{1}}{(\alpha \mu_B B)^{3/2}}$
-- the case for \ch{H3LiIr2O6} \cite{Kitagawa2018}. 

For $\eta=0.9$ and $\gamma=1.8$, we get $\frac{C_m}{T} \propto |T|^{-0.9}$ and $B^{0.9}\left(\frac{C_m}{T}\right) \propto \left(\frac{T}{B}\right)^{1.8}$ -- which is the case for our compound HLRO and hence our model DOE: $D(E,0) = \Gamma |E|^{-0.9}$  and $D(E,B\ne0)= \Gamma \frac{|E|^{1.8}}{(\alpha \mu_B B)^{2.7}}$. 

Alternative approach in Literature\cite{Andrade2022},

Bulk magnetic susceptibility($\chi$) at low-$T$:
\begin{equation}
	\chi \sim T^{-\eta}
	\label{(andrade1)}
\end{equation}
where $\eta$ is an exponent.

Heat capacity at $B=0$;
\begin{equation}
	C_m/T \sim T^{-\eta}
	\label{(andrade2)}
\end{equation}

Heat capacity at $B\ne0$;
\begin{equation}
	C_m \sim B^{-3\eta}T^{1+2\eta}
	\label{andrade3}
\end{equation}

In HLRO, $\chi \sim T^{-0.9}$ and 	$C_m/T \sim T^{-0.9}$. From equation \ref{andrade3}, we get $C_m \sim B^{-2.7}T^{2.8}$. That implies $B^{0.9}\left(\frac{C_m}{T}\right) \propto \left(\frac{T}{B}\right)^{1.8}$ -- which is the case for our compound HLRO.

Our model DOE and alternative approach in literature\cite{Andrade2022}, are in consistent with our experimental observed scaling.







\section{Scaling in $M$, $C_{m}$ and 1/T$_{1}$}

\begin{figure}[!h]
	\centering
	\includegraphics[width=0.9\linewidth]{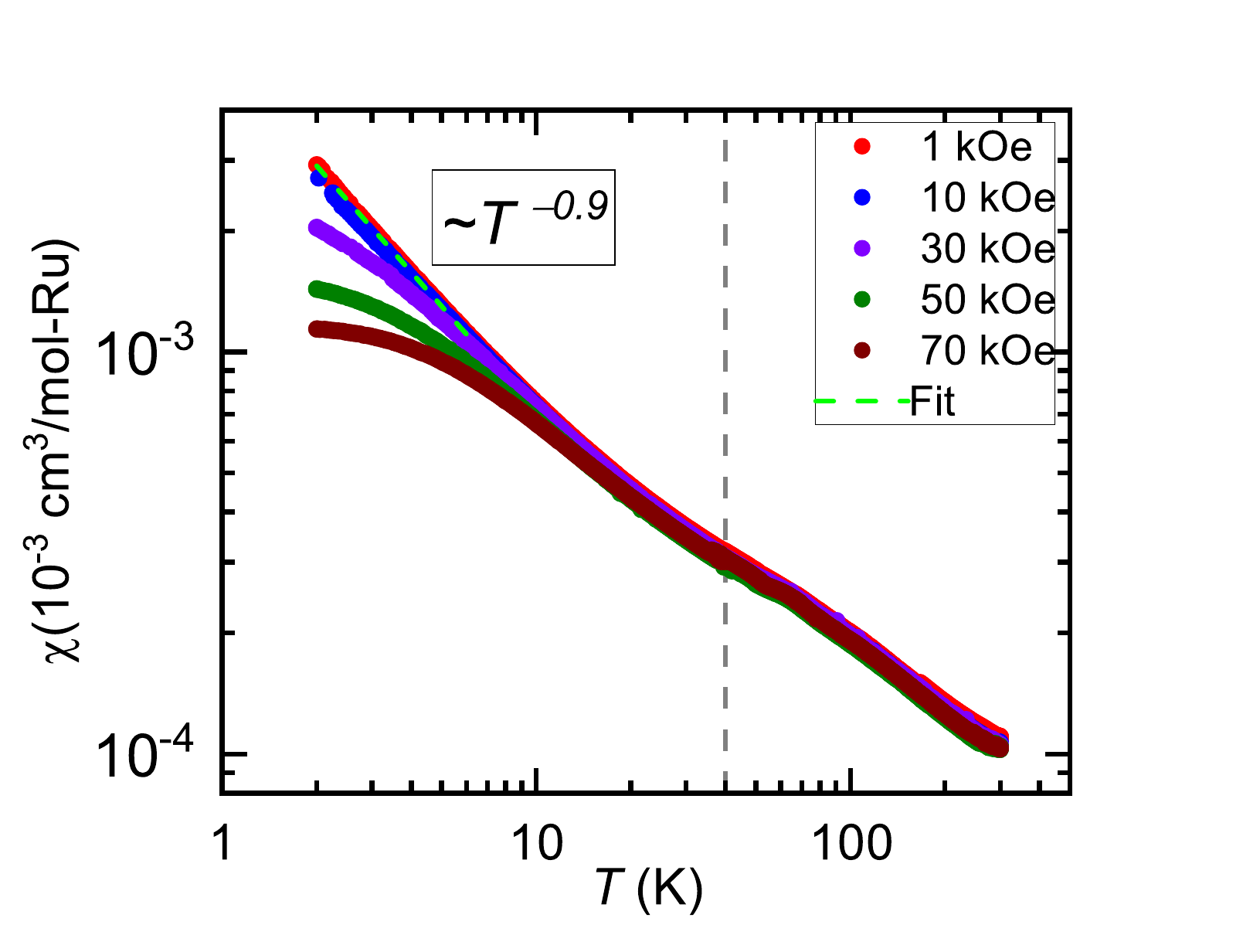}\caption{Magnetic susceptibility, $\chi$ as function of temperature for various fields in the range 1-70 kOe. The green dot line indicates low-$T$ upturn of $ \chi_{bulk}$ ($ \sim T^{-0.9}$).}
	\label{chi_field}
\end{figure}

\begin{figure}[!h]
	\centering
	\includegraphics[width=0.9\linewidth]{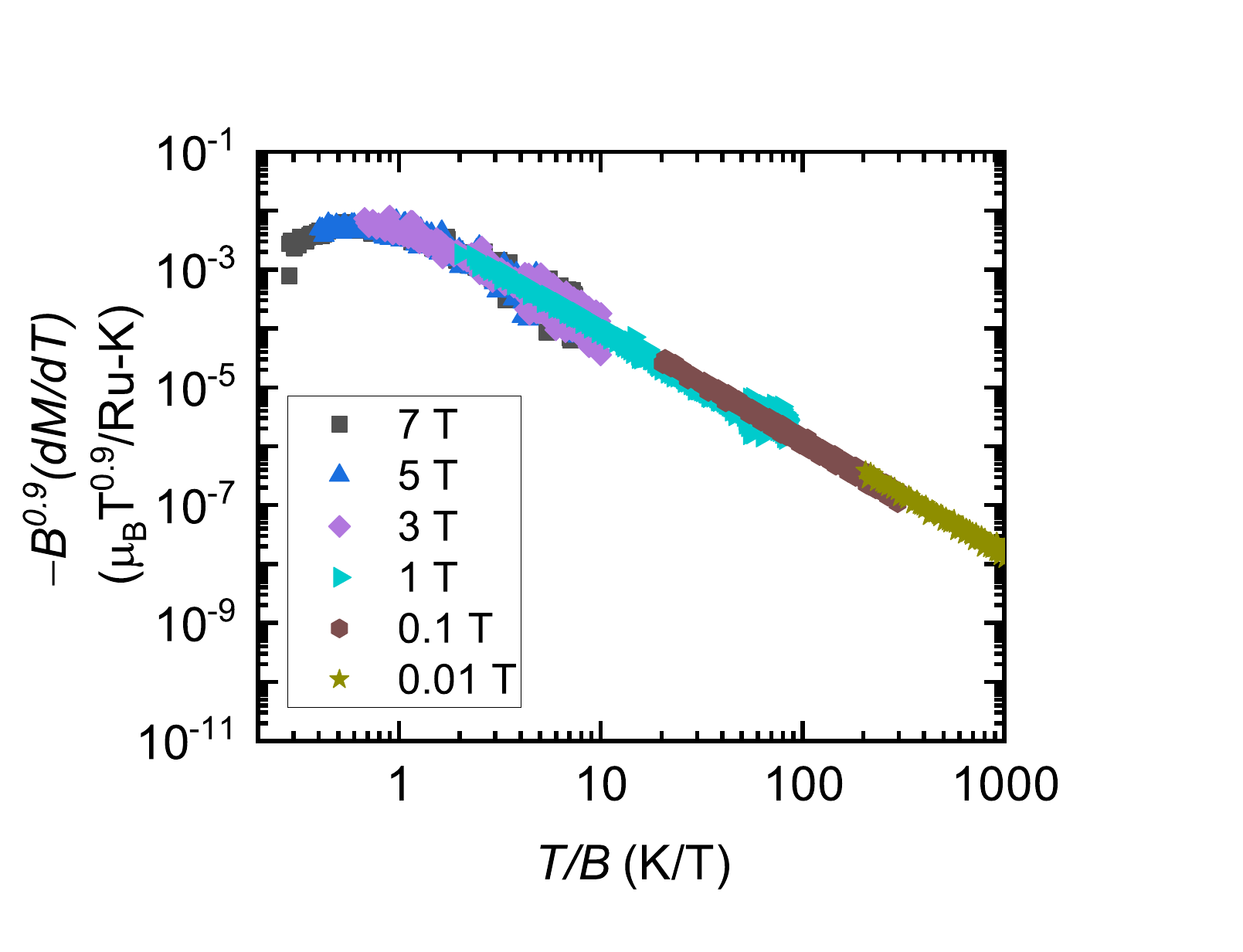}
	\caption{A scaling of $B^{0.9}(dM/dT)$ with $T/B$}
	\label{MT_scaling}
\end{figure}

\begin{figure}[!h]
	\centering
	\includegraphics[width=0.9\linewidth]{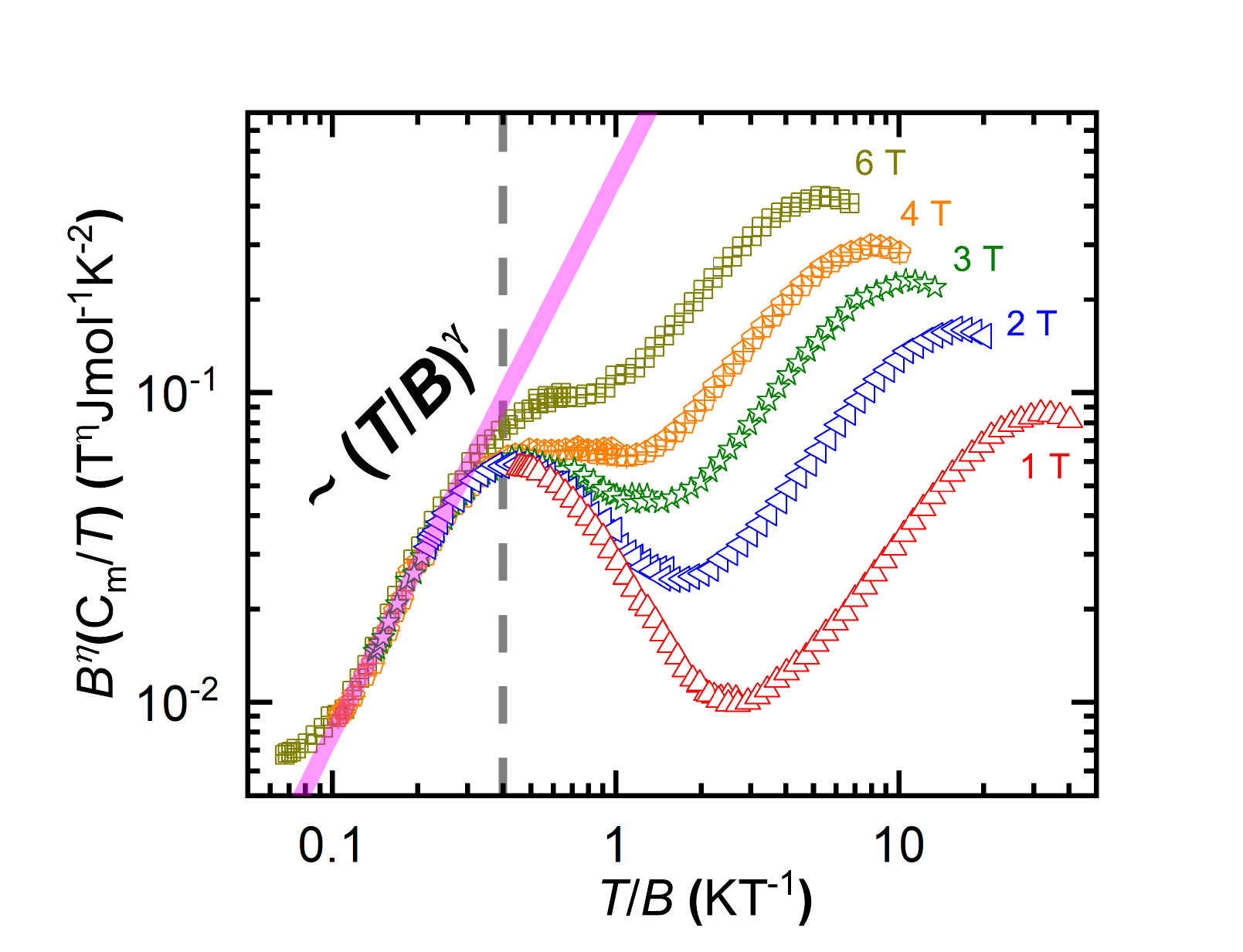}
	\caption{A scaling/data-collapse of $B^\eta(C_{m}/T)$ with $(T/B)^\gamma$ is seen if we take $\eta=0.9$ and $\gamma=1.8$.}
	\label{Cm_scaling}
\end{figure}

At low-$T$, bulk susceptibility (magnetization measurement) behaves like conventional Curie-Weiss ($\chi_{bulk} \propto T^{-1}$) due to magnetic defects (of extrinsic) where intrinsic susceptibility (NMR Shift) is independent of temperature, a reminiscent of QSL. However, $\chi_{bulk} \propto T^{-\eta}$ ($0<\eta<1$) is due to magnetic defects of intrinsic origin like vacancy, quasi-vacancy, bond-randomness and so on. $\chi_{bulk} \propto T^{-0.5}$ was found in \ch{H3LiIr2O6} \cite{Kitagawa2018} and different values of n=0.68 is reported \cite{Yong2023} (sample dependency). Figure \ref{chi_field} shows magnetic susceptibility, $\chi$ as function of temperature for various fields in the range 1-70 kOe. The green dot line indicates low-$T$ upturn of $\chi_{bulk}$ ($\sim T^{-0.9}$). With an application of magnetic field, it suppressed and become almost $T$-independent at 70 kOe. The behavior is qualitatively similar like \ch{H3LiIr2O6} \cite{Kitagawa2018} but exponent is different in our case. We speculate presence of magnetic defects in the form of vacancy and quasi-vacancy. The effects of vancancy and quasi-vacancy has been seen in zero field $C_{m}$ ($\propto T^{-0.9}$) as well (theoretical simulation can be found in literature \cite{Perkins2021}). We found a scaling of $B^{0.9}(dM/dT)$ with $T/B$ (Figure \ref{MT_scaling}). Magnetic heat capacity is also field dependent and shows a scaling of $B^\eta(C_{m}/T)$ with $T/B$ and exponent, $\gamma=1.8$ and $\eta=0.9$. This arises from field dependent density-of-states as seen in \ch{H3LiIr2O6} \cite{Kitagawa2018}. We found that the phenomenological model $D(E,B) = \Gamma \frac{|E|^{1.8}}{(\alpha \mu_B B)^{2.7}}$ explains well our experimental results of $C_{m}$.       

\begin{figure}[!h]
	\centering
	\includegraphics[width=0.9\linewidth]{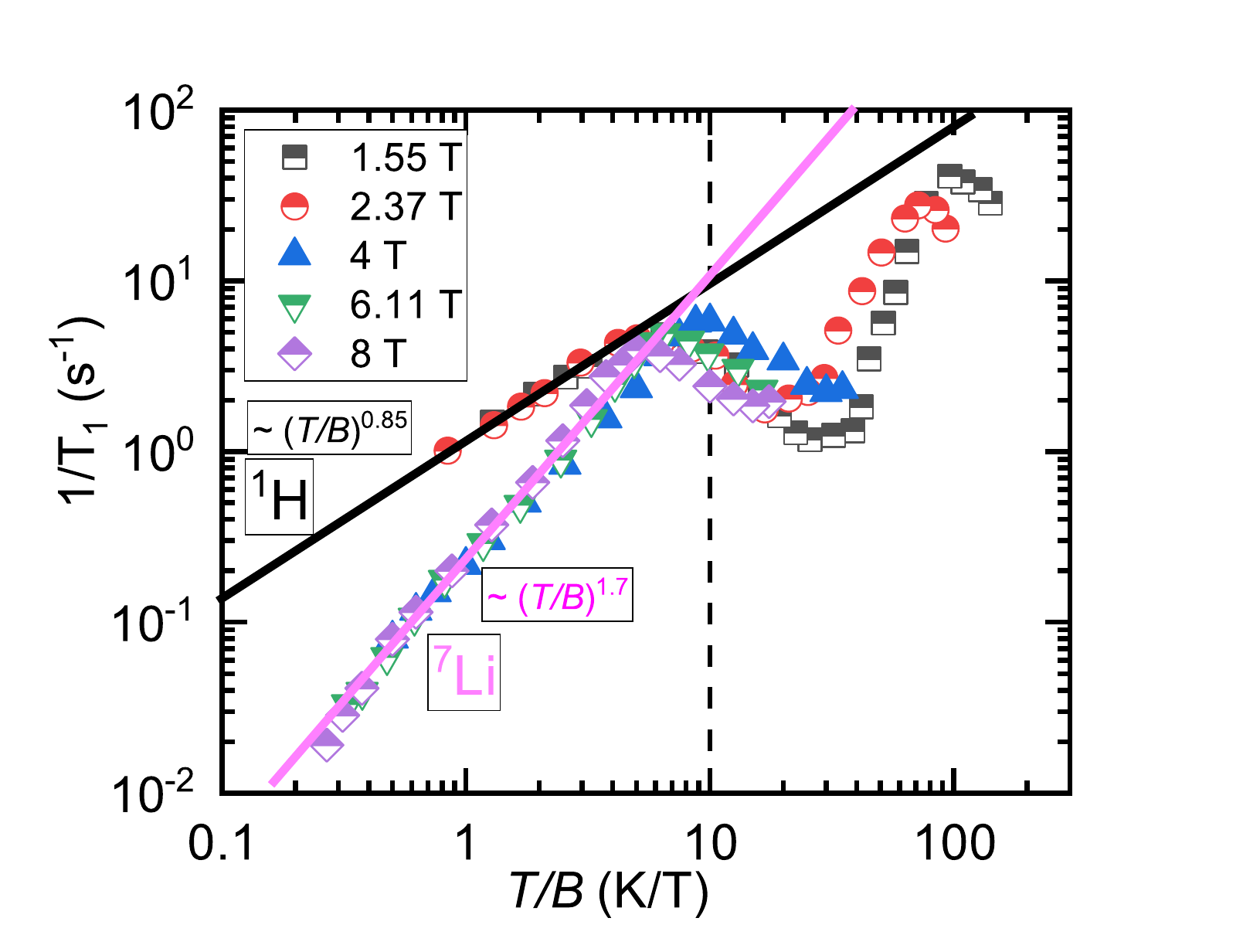}
	\caption{The temperature-field scaling of $^{7}$Li-Spin-lattice
		relaxation rate, 1/T$_{1}$ with exponent $\sim$ 1.7 and $^{1}$H-Spinlattice relaxation rate, 1/T$_{1}$ with exponent $\sim$ 0.85 for \ch{H_{5.9}Li_{0.1}Ru2O6}. The data for $^{1}$H are multiplied by 13.24 to overlap with those for $^{7}$Li.}
	\label{T1_scaling}
\end{figure}

The susceptibility from defect regions is expected to have a power law increase with decreasing temperature. While this shows up as an upturn with decreasing temperature
in the bulk susceptibility, the NMR line shape should have a contribution (from defects) which has a power law increase in the shift with decreasing temperature. At low temperature, the defect contribution should then be well separated from the main resonance line. Consequently, in our T$_{1}$ measurements, where the transmitter frequency is centered on the main line, we are largely not sensitive to the defect contribution (which is out of our spectral width). We observed field-dependent 1/T$_{1}$ and a temperature-field scaling of $^{7}$Li-Spin-lattice
relaxation rate, 1/T$_{1}$ with exponent $\sim$ 1.7 and $^{1}$H-Spinlattice relaxation rate, 1/T$_{1}$ with exponent $\sim$ 0.85 for \ch{H_{5.9}Li_{0.1}Ru2O6}. This also arises from field-dependent density of states but not same as $D(E,B) = \Gamma \frac{|E|^{1.8}}{(\alpha \mu_B B)^{2.7}}$ (valid for $C_{m}$ since heat capacity is a bulk probe and captures excitations from defects as well). Intrinsic magnetic defects (vacancy and quasi-vacancy) does effect the ground states. The picture in HLRO is that ground state has bound Z$_{2}$ flux in low fields in presence of vacancies. The zero-flux KQSL is obtained upon the application field. The power law scaling is consistent with gapless excitations which we ascribed to the zero-flux KQSL and weak localization of Majorana fermions.

\bibliographystyle{plain}
\bibliography{citation}